\documentclass[preprint,12pt]{elsarticle}
\biboptions{sort&compress}

\usepackage{amssymb}
\usepackage{amsmath}
\usepackage{placeins}
\usepackage{xcolor}
\usepackage{multirow}
\usepackage{soul}
\usepackage{caption}
\usepackage{subcaption}
\usepackage{listings}
\definecolor{codegreen}{rgb}{0,0.6,0}
\definecolor{codegray}{rgb}{0.5,0.5,0.5}
\definecolor{codepurple}{rgb}{0.58,0,0.82}
\definecolor{backcolour}{rgb}{0.95,0.95,0.92}

\lstdefinestyle{mystyle}{
    backgroundcolor=\color{backcolour},   
    commentstyle=\color{codegreen},
    keywordstyle=\color{magenta},
    numberstyle=\tiny\color{codegray},
    stringstyle=\color{codepurple},
    basicstyle=\ttfamily,
    breakatwhitespace=false,         
    breaklines=true,                 
    captionpos=t,                    
    keepspaces=true,                 
    numbers=left,                    
    numbersep=5pt,                  
    showspaces=false,                
    showstringspaces=false,
    showtabs=false,                  
    tabsize=2
}

\newcommand{\R}{\textsuperscript{\textregistered}}

\journal{Additive Manufacturing}

\begin{document}

\begin{frontmatter}



\title{Pressure drop non-linearities in material extrusion additive manufacturing: a novel approach for pressure monitoring and numerical modeling}


\author[inst1]{Sietse de Vries \corref{cor2}}
\author[inst2,inst3,inst4]{Tom\'as Schuller \corref{cor2}}
\author[inst4,inst5]{Francisco J. Galindo-Rosales }
\author[inst6]{Paola Fanzio \corref{cor1}}

\cortext[cor1]{Corresponding author}
\cortext[cor2]{These authors contributed equally to this work.}

\affiliation[inst1]{organization={Ultimaker B.V.},
            addressline={Watermolenweg 2}, 
            city={Geldermalsen},
            postcode={4191 PN},
            country={The Netherlands}}
            
\affiliation[inst2]{organization={Institute of Science and Innovation in Mechanical and Industrial Engineering (INEGI)},
            addressline={Rua Dr. Roberto Frias, 400}, 
            city={Porto},
            postcode={4200-465},
            country={Portugal}}

\affiliation[inst3]{organization={Transport Phenomena Research Center (CEFT), Mechanical Engineering Department},
            addressline={Faculty of Engineering of the University of Porto, Rua Dr. Roberto Frias s/n}, 
            city={Porto},
            postcode={4200-465}, 
            country={Portugal}}

\affiliation[inst4]{organization={ALiCE—Associate Laboratory in Chemical Engineering},
            addressline={Faculty of Engineering of the University of Porto, Rua Dr. Roberto Frias s/n}, 
            city={Porto},
            postcode={4200-465}, 
            country={Portugal}}

\affiliation[inst5]{organization={Transport Phenomena Research Center (CEFT), Chemical Engineering Department},
            addressline={Faculty of Engineering of the University of Porto, Rua Dr. Roberto Frias s/n}, 
            city={Porto},
            postcode={4200-465}, 
            state={},
            country={Portugal}}
\affiliation[inst6]{organization={Department of Precision and Microsystems Engineering (PME), Faculty of Mechanical, Maritime and Materials Engineering (3mE)},
            addressline={Delft University of Technology), Mekelweg 2}, 
            city={Delft},
            postcode={2628 CD}, 
            country={The Netherlands}}
            
\begin{abstract}
Fused Filament Fabrication is an additive manufacturing technique in which molten thermoplastic polymers are extruded through a nozzle. Therefore, the interplay between the viscoelastic nature of the polymer melt, temperature, printing conditions and nozzle shape may lead to inconsistent extrusion. 
To improve the extrusion control and optimize the print-head performance, a better understanding of the flow process of the polymer melt both in the nozzle and the liquefier is needed. However, several challenges need to be overcome due to the complexity of gathering experimental data on the melt pressure in the nozzle and the lack of numerical models able to capture the full rheology of the molten polymer. 
This research introduces an innovative approach for monitoring the pressure within a material extrusion 3D printer's nozzle. This method involves utilizing a pin in direct contact with the molten material, which then transmits the applied force from the material to an externally mounted load cell. The setup provides reliable, repeatable pressure data in steady-state conditions for two nozzle geometries and at different extrusion flows and temperatures. 
Moreover, the Giesekus model enabled capturing the viscoelastic rheometric features of the melt, and the numerical predictions have been compared with the experimental data. Results show that the numerical model accurately describes the flow conditions in the nozzle and allows the estimation of the behaviour of the melt in the liquefier zone, the area of the print-head where the filament is molten. It could be concluded that the backflow, which is the backward flow of the molten polymer in the gap between the filament and the liquefier towards the cold end, caused significant non-linearities in the total pressure drop measured in the feeders, which were related to normal forces induced by shear in that region.

\end{abstract}

\begin{graphicalabstract}
\includegraphics[width=1.3\textwidth]{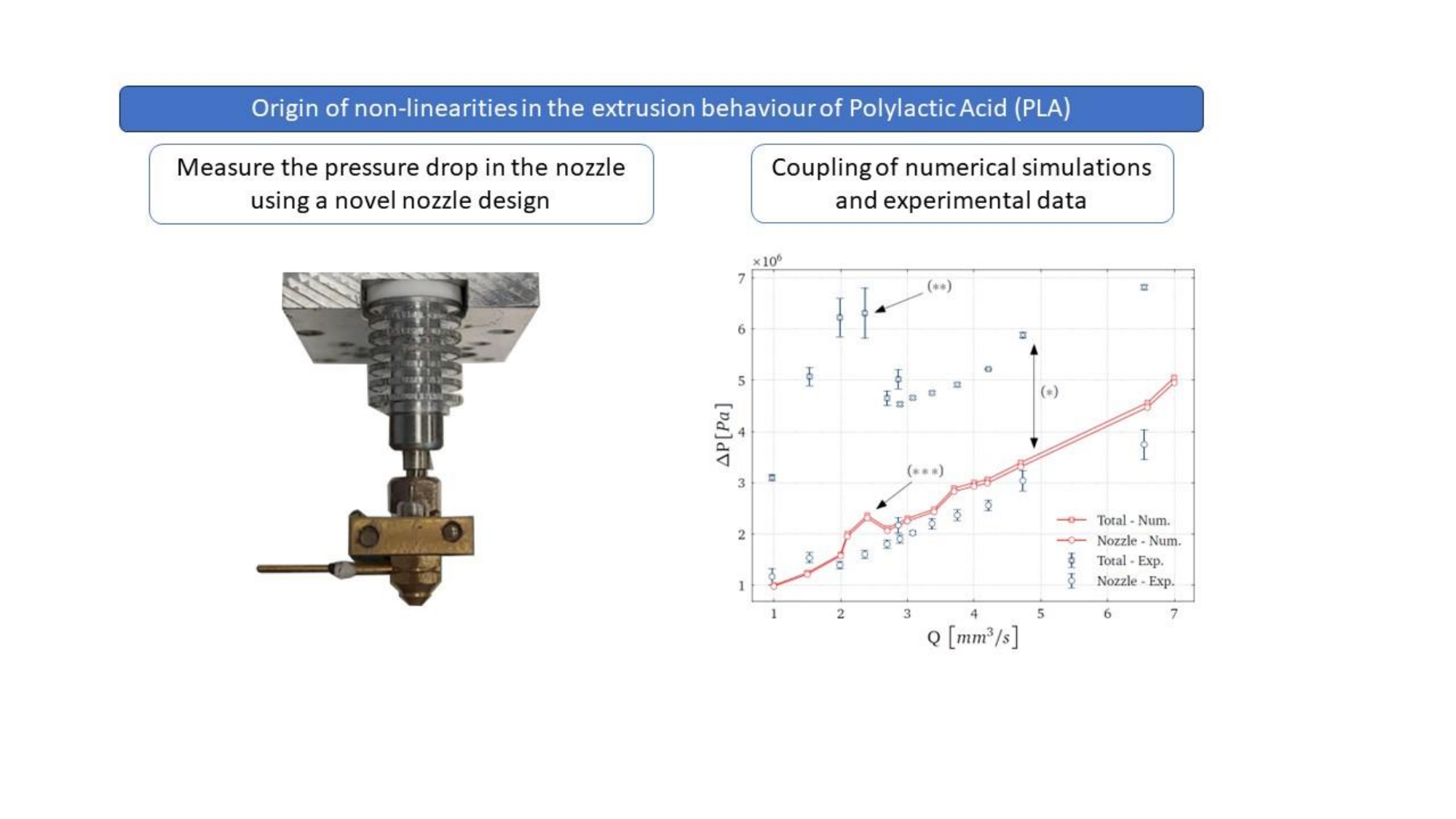}
\end{graphicalabstract}

\begin{highlights}
\item A novel nozzle design for measuring the pressure drop in the nozzle;
\item Coupling of numerical simulations and experimental data enables the explanation of extrusion behaviour in Polylactic Acid (PLA);
\item Elastic instabilities in the melt flow within the nozzle are responsible for non-linearities in nozzle pressure drop;
\item The liquefier backflow phenomenon appears to contribute significantly to the extrusion pressure drop.

\end{highlights}

\begin{keyword}
3D printing \sep Fused filament fabrication  \sep Polylactic acid (PLA) \sep Numerical simulations \sep Elastic instabilities  
\PACS 0000 \sep 1111
\MSC 0000 \sep 1111
\end{keyword}

\end{frontmatter}


\FloatBarrier
\section{Introduction}
\label{sec:sample1}
\begin{figure}[ht!]
\centering
    \includegraphics[width=\textwidth]{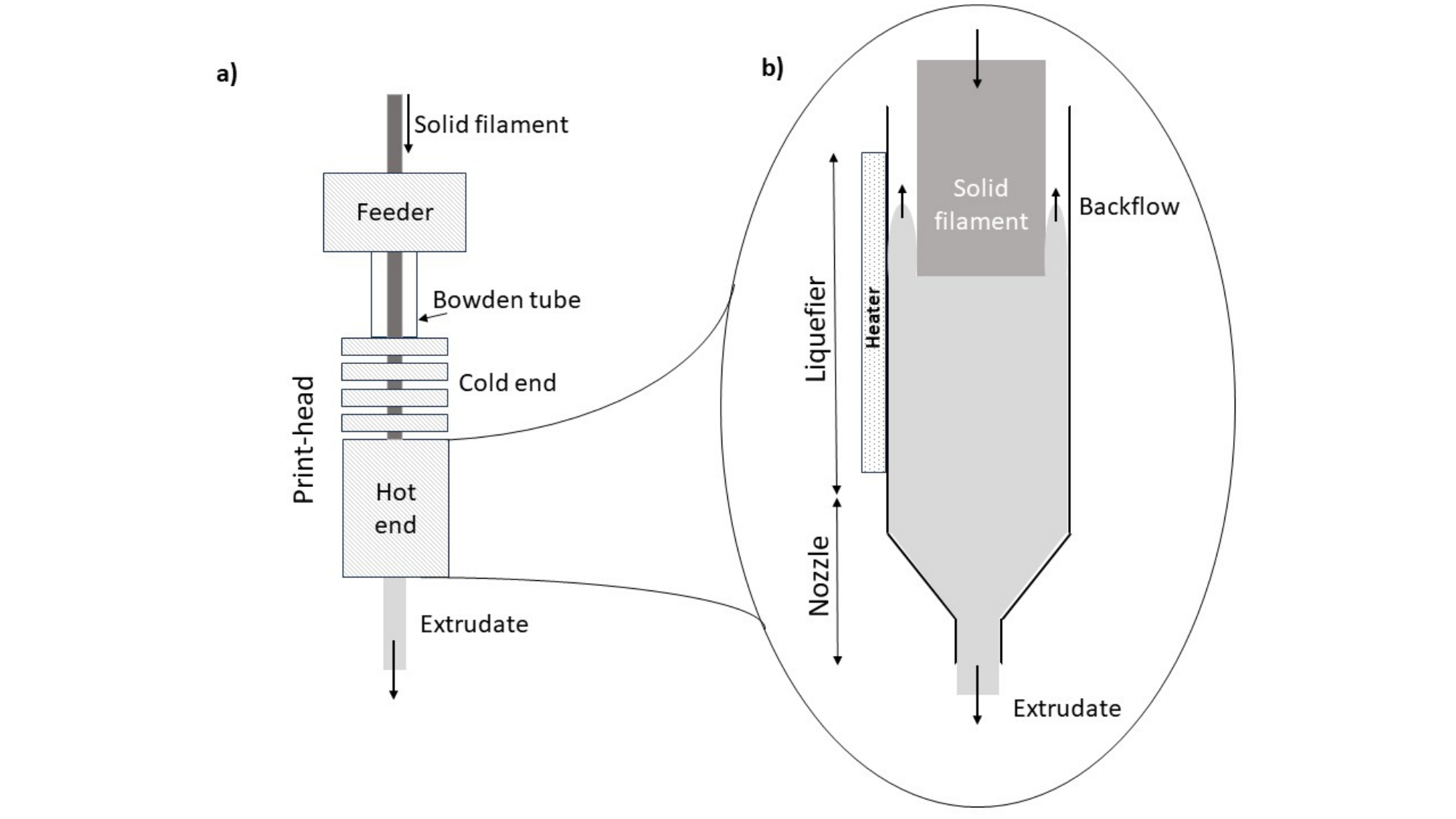}
    \caption{a) Scheme of a extruder head; b) Scheme of the hot end region of the print-head.}
    \label{fig:1}
\end{figure}

Material extrusion is the most widely used additive manufacturing technique due to its user-friendly workflow, relatively low cost and broad material availability. The working principle is relatively simple (Figure~\ref{fig:1}a): a solid polymeric filament, confined in a low-friction Bowden tube, is moved towards the print-head using a feeder (rotating rollers actuated by a motor). In the print-head the filament enters the liquefier, where it is melted, and the melt is extruded through a nozzle. The final 3D object is manufactured in layers, each being created by the deposition of polymer in adjacent lines by means of the movement of a print-head \cite{turner}.

Despite the academic research effort and the industrial development performed, several challenges still need to be addressed in order to increase the performance of material extrusion printers. Print quality defects, affecting dimensional accuracy, mechanical properties and visual quality, are typically linked to inconsistent extrusion (under- or over-extrusion) \cite{turner2}.

Flow control strategies are implemented to keep the speed of the solid filament entering the print-head constant. However, pressure variations inside the print-head are not considered, leading to poor control of the actual melt flow exiting the nozzle.

Such pressure variations are due to the complexity of the extrusion process itself. In fact, the melt inside the print-head behaves as a viscoelastic fluid, subjected to both extensional and shear flow conditions and temperature gradients \cite{Das}. Moreover, it is essential to note that the filament supplies the material for extrusion and also pushes the melt to the nozzle (Figure~\ref{fig:1}b). An annular gap between the solid filament entering the print-head and the wall of the liquefier is then present, allowing an unwanted backflow of the polymer melt toward the cold end zone of the print-head, in the opposite direction of the extrudate flow. In addition, the strong temperature gradient in the liquefier and its dependency on the volumetric flow rate influence the melt pressure inside the print-head, affecting both the actual volumetric melt flow exiting the nozzle and its temperature.

Obtaining a comprehensive grasp of the extrusion process is essential for the development of advanced flow control strategies. In this context, numerical models have been developed to describe the extrusion process, capturing some of the phenomena described above. 

Bellini et al. \cite{bellini} proposed one of the first models, estimating analytically the nozzle pressure drop by solving the momentum flow balance. However, in this work, several assumptions have been made, such as the use of the Arrhenius temperature-viscosity relationship and the use of the power law model, which is valid only for viscosity values in a limited range of shear rates. A more comprehensive extrusion model has been developed by T.A. Osswald et al. \cite{osswald} by refining the description of the melting process, including all process parameters (initial filament temperature, heater temperature, applied force, among others$\ldots$), geometrical parameters and shear viscosity data of the melt.

Moretti et al. \cite{Moretti2020} have recently proposed an evolution of such a model, where the print-head has been sensorized with a load cell added between the print-head and the feeder for in-process simulation and real-time monitoring of the total extrusion pressure drop. The scope of their work was to create a digital twin system capable of predicting flow rate, extrudate temperature and compression stress on the filament. However, such a model was still unable to capture the full rheological characteristic of the polymeric melt. In particular, it did not take into account the enhanced elasticity of a viscoelastic material, which affects the melt flow in the nozzle \cite{rooooooooy}.

Gilmer et al. \cite{gilmer} proposed the introduction of a Flow Identification Number (FIN) to predict the material’s propensity to backflow, which was subsequently reformulated by Mackay \cite{mackay1}. The FIN was defined as:

\begin{equation}
FIN=\frac{\frac{\Delta P_{liq}}{H} (D-D_f)^2}{\eta V_f }
\end{equation}

Where $\Delta P_{liq}$ is the pressure drop in the liquefier, H is the liquefier length, $\eta$ is the melt viscosity (in the Newtonian plateau), $V_f$ is the filament velocity, D is the diameter of the liquefier and $D_f$ is the filament diameter. The authors claim that if the FIN is greater than 1, the length of the backflow will exceed the length of the liquefier flowing in the cold end of the print-head, possibly causing an extrusion failure. The FIN calculation is based on a balance between drag flow (produced by the filament entering the liquefier) and the pressure flow (produced by the pressure drop) but assumes a Newtonian viscosity and a uniform temperature distribution.

Despite all the effort, a complete understanding of the melting and extrusion process is still lacking. Experimental data are of great value to validate current models, and, in particular, measurements of pressure in the nozzle are needed to discriminate between the physical processes happening in the liquefier and in the nozzle.

From an experimental point of view, Fang Peng et al. \cite{peng} experimentally quantified the temperature evolution and the flow behaviour by using die markers within the filament, which would enable to visualize the flow behaviour by looking at the pigment distribution in extruded filaments and in the filament inside the nozzle.

Anderegg et al. \cite{anderegg} pioneered a novel nozzle configuration featuring a piezoresistive pressure transducer, capable of simultaneously gauging temperature and pressure within the flow field. This innovation was aimed at facilitating real-time process monitoring for quality control purposes. Initially, their setup faced challenges due to its larger heating block, resulting in difficulties maintaining a stable nozzle temperature. However, they managed to resolve this issue by revising the PID control loop for small flow rates. This adjustment led to improved performance, with pressure predictions based on Bellini's model closely approximating about 73\% of the experimental values.

Coogan and Kazmer \cite{coogan_kazmer} developed an in-line rheometer to measure material viscosity from pressure data. They used a custom nozzle, a load-transfer column, a thermocouple, clamps, and a load cell. They obtained very precise viscosity values. However, their design had drawbacks, such as nozzle wear and large size. They also experienced filament leakage, which required them to clean the nozzle and reset the set-up after long prints. These problems could be solved by reducing leakage with better machining and using the same materials for different parts to match thermal expansion.

In our previous paper \cite{schuller_add_ma}, we simulated the extrusion process of viscoelastic polymers in material extrusion, including both shear and extensional viscosity and showing the formation of upstream vortices. Experimental extrusion pressure drop data was acquired using a print-head equipped with a load cell. Numerical results have been compared with experimental data, and an improved evaluation of the FIN number was proposed. However, experimental data on the pressure conditions inside the liquefier were still missing.

This paper aims to solve such an issue by comparing the extrusion pressure data measured at two locations in the print-head. A new nozzle design is proposed, which includes a force sensor in contact with the melt and a second force sensor placed between the feeder and print-head. Numerical simulations have been performed, taking into account the viscoelastic nature of the polymer melt as well. Liquefier and nozzle pressure drops have been measured separately and compared with numerical results, enabling, for the first time, the discrimination of the phenomena happening in the liquefier and nozzle areas. 

\FloatBarrier
\section{Material and Experimental Methods}

\subsection{Materials}
All the extrusion experiments have been performed using a 2.85 $\pm$ 0.01 mm diameter filament spool of transparent polylactic acid (PLA) from UltiMaker. 

\subsection{Sensorized nozzle design}
\begin{figure}[ht!]
\centering
    \includegraphics[width=0.9\textwidth]{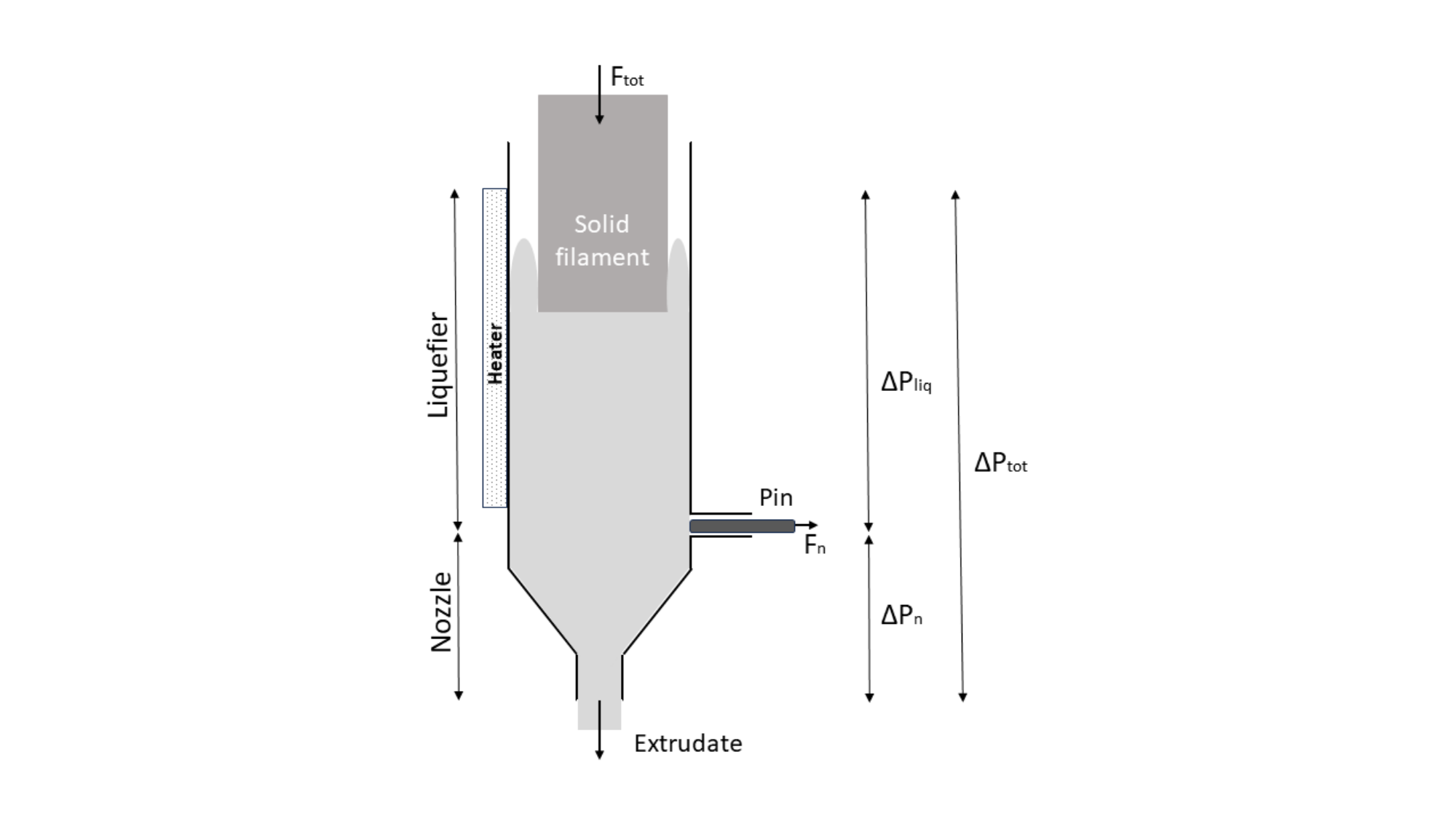}
    \caption{Schematic representation of the sensorized nozzle concept.}
    \label{fig:2}
\end{figure}

A new nozzle design is proposed to enable the measurement of the pressure drop in the nozzle ($\Delta P_n $) by using a pin in direct contact with the melt through a hole on the side of the nozzle. This concept is explained in Figure~\ref{fig:2}, where a schematic of the sensorized nozzle is presented. The melt flowing in the nozzle exerts a force on the pin ($F_{n}$), measured with an external load cell.

\begin{figure}[ht!]
\centering
    \includegraphics[width=\textwidth]{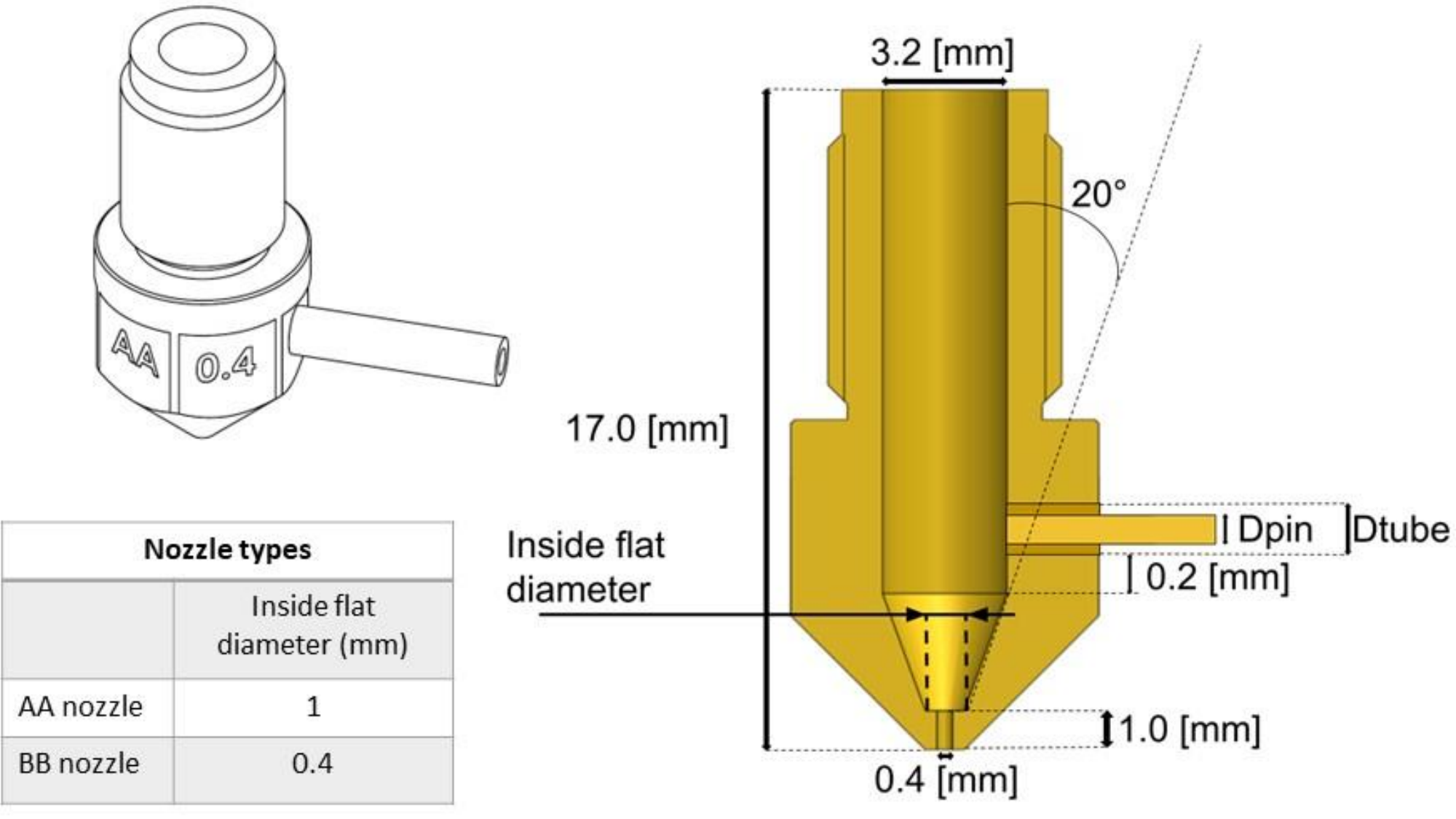}
    \caption{Dimensions and type of sensorized nozzles.} 
    \label{fig:4}
\end{figure}

Figure~\ref{fig:4} shows a mechanical drawing of the sensorized nozzle. An aperture was created in the sidewall of a standard UltiMaker brass nozzle. To prevent any potential issues such as locking and buckling of the pin, a brass guiding tube was press-fit into the nozzle to act as a guiding mechanism for the pin. The pin was then inserted into this guiding tube. A sliding fit H7/g6 was chosen, as it is well-suited for applications with very small clearances, ensuring precise alignment of the shaft. To avoid thermal expansion discrepancies, the pin was manufactured from the same material as the nozzle, which is brass.
Prior to testing, the adapted nozzle and the pin were imaged with a Leica DVM6 Digital Microscope (see Supplementary Information). From the images, the diameters of the pin $D_{pin} =$ 1 mm, the diameter of guidance tube $D_{tube} =$ 1 mm and the nozzle die diameter $D_{die} =$ 0.4 mm were calculated. The length of the liquefier (from the guidance tube to the cold end) is 11.8 mm.
Two types of nozzle shapes have been evaluated: nozzle shape AA exhibits a flat area between the tapered region and the die (1mm in diameter), and nozzle shape BB has no flat area. Dimensions of the tested nozzles are summarized in Figure~\ref{fig:4}.

\subsection{System design}
\begin{figure}[ht!]
\centering
    \includegraphics[width=\textwidth]{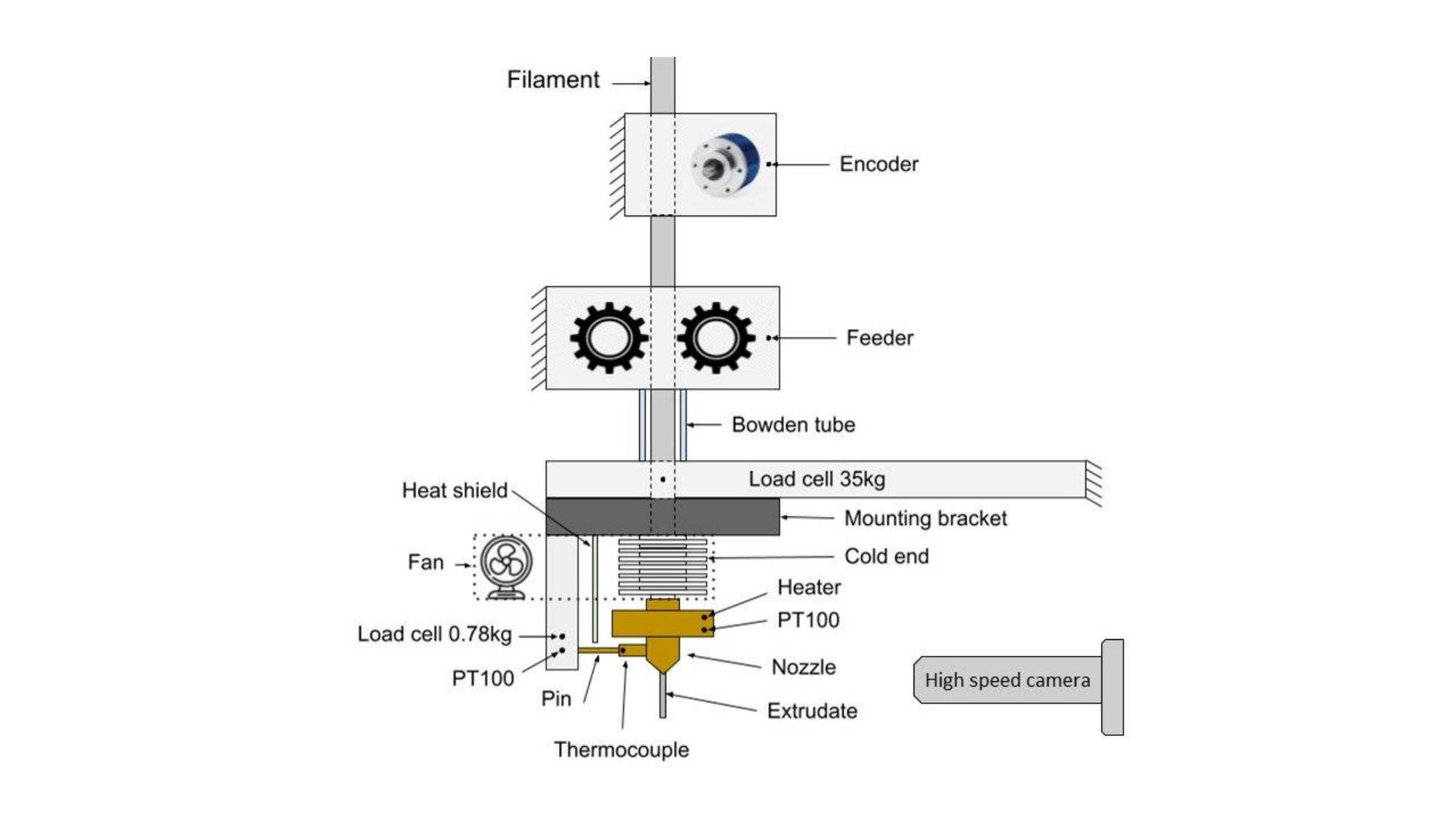}
    \caption{System of sensors, devices, and connections for extrusion measurement setup.}
    \label{fig:5}
\end{figure}

Figure~\ref{fig:5} shows a schematic of the complete test setup. Following the standard print-head assembly in UltiMaker printers: the cold end is mounted to the sensorized nozzle, which is heated through an heater block.\\
A 25 watt heater cartridge coupled with a platinum 100 ohm (PT100) temperature sensor control the temperature of the heater block (PID calibrated with the following values: $kp =$ 0.04, $ki =$ 20.0 and $kd =$ 10.0).

A Conatex TM12K02GG2 thermocouple is glued to the end of the guidance tube using a temperature-resistant glue, Permasol Sauereisen No. 31 A/B adhesive, which also serves as a good heat conductor. The pin in contact with the melt transfers the force to a 0.78 kg Wheatstone bridge load cell ($F_{n}$, in Figure~\ref{fig:2}).
The temperature of the load cell is continuously monitored with a PT100. To prevent overheating of the 0.78 kg load cell, an aluminum foil heat shield is in place. Cooling for both the load cell's cold end and the load cell itself is provided by a Sunon axial fan. Additionaly a small amount of lubricant is applied between the pin and the load cell. \\
The force applied by the feeder to push the filament ($F_{tot}$, in Figure~\ref{fig:2}) is measured by a second load cell (35 kg Wheatstone bridge). This load-cell is positioned between the feeder and the print-head. \\
All load cells, the fan, the heat shield, and the cold end, are securely affixed to a custom-made aluminum mounting bracket.\\
In this experiment, the feeder comprises an MS16HS7P4070 stepper motor responsible for advancing the filament into the liquefier. To monitor the filament's speed as it enters the liquefier, an ERN 1023 rotary encoder is employed. The filament is guided into the liquefier through a Bowden tube.\\
Data from the sensors are logged at a rate of 1 kHz using Beckhoff data acquisition modules connected to real-time TwinCAT software through EtherCAT. All data points are synchronized on the same timestamp.
The experiment was conducted as a proof-of-concept in a static environment, with no movement of the print-head. This setup differs from standard printing condition since, usually, the print-head moves in the XY plane to deposit lines of molten polymer. 
Additionally, the extrusion is carried out in the open air rather than on a build plate, diverging from the typical printing setup. 

\subsection{Calibration}
The calibration of the feeder and rotary encoder involved moving 20.0 mm and 100 mm of filament, measuring the encoder values, and adjusting the encoder count per unit distance in the encoder configuration file accordingly. The same calibration process was performed to calibrate the feeder steps. This calibration process was repeated until the discrepancies in encoder and feeder values were minimized to within 0.01\%.
A Sauter FH-S force gauge mounted on a worm wheel that could be manually adjusted was used to calibrate both load cells. The deflection of the 0.78 kg load cell was measured with a caliper attached to the same worm wheel. A force of 7.0 N applied on the load cell resulted in a deflection of 0.07 mm (the force measured at the empty nozzle was zero). Two more calibrations were done by applying the force at different points on the load cell. The measurement error was larger than the difference between the points, indicating that the pin position relative to the load cell had a negligible effect. 
A thermocouple, previously calibrated using a RS PRO temperature calibrator, was then used to calibrate both PT100s immersed in a polymer melt at 200 \textdegree C. Standard UltiMaker PID settings were used for temperature control.
The temperature rose from 20 \textdegree C to 200 \textdegree C with a 1 \textdegree C overshoot and stabilized in $\approx$ 2 seconds.

\subsection{Data acquisition}
At the start of the experiment, the print-head received the filament at a constant feed rate and temperature. Purging the nozzle was performed for 30 minutes before each experiment. After purging, the system was idled and scouted for failures, and the data was checked for anomalies. \\
Extrusion measurements were performed at three temperatures ($195\,^{\circ}$C, $205\,^{\circ}$C, $215\,^{\circ}$C) by varying the input filament volumetric flow rate (also called flow in the following paragraphs) from 1 mm$^3$/s to 7 mm$^3$/s.
Constant extrusion was recorded for 10 minutes and after every test, the data was stored as a single hierarchical data format 5 file (hdf5). A ten-second pause separates the tests from each other. 
For every temperature, the test was repeated twice.

The average force is calculated for each test with a constant temperature and feed rate. Data acquired in the first 150 s was not used in the calculation to let the extrusion stabilize and reach a steady state. 

\subsection{Die swelling measurement and analysis}
Die swelling measurements have been performed using a high-speed camera (PL-D725MU-T) from Pixelink, equipped with a standard adapter tube 1X (1-6515) and a converter lens 1X (1-60640) by UltraZoom. The camera has been placed in front of the nozzle, as shown in Figure~\ref{fig:5}. A high-lumen lamp has been used to improve the lighting conditions.
After reaching steady-state conditions, three images of the filament extruded from the nozzle were acquired for each volumetric flow rate. Calibration of the pixel dimension has been performed using a needle with a known diameter.
Despite gravity not being a significant factor in the die swell phenomenon \cite{Tanner1970}, the extruded filament was cut with an initial length of $\approx$ 5 cm before the image acquisition.
Images have been analysed using ImageJ \cite{imagej}, and the extrudate diameter has been evaluated at a fixed distance of 500 µm from the nozzle die. 

\subsection{Rheological characterization}

The viscosity measurement under constant shear conditions was conducted using an ARES G2 rotational rheometer at a temperature of 200~$^\circ$C. This involved using parallel plates with a diameter of 12 mm and a 1 mm gap. For the shear rheological tests, specimens were shaped into disks with approximately 12 mm in diameter and 1.5 mm in thickness, achieved through a hot compression moulding process. To prepare PLA specimens, they were placed in a well-ventilated oven at 40~$^\circ$C for a minimum of 4 hours. The compression moulding was executed at 200~$^\circ$C and subsequently cooled using a water circulation system.

The assessment of elongational viscosity was conducted using a rotational rheometer (ARES G2), equipped with a TA Instruments Extensional Viscosity Fixture (EVF)~\cite{EVFpatent,EVFTAinstr1,EVFTAinstr2}, operating at a temperature of 200~$^\circ$C. Various strain rates, including 0.1 s$^{-1}$, 0.3 s$^{-1}$, 1 s$^{-1}$, and 3 s$^{-1}$, were applied. Specimens with roughly 8-10 mm in width, 1 mm in thickness, and 18 mm in length were prepared by die-cutting from plates obtained through compression moulding. Cooling was achieved through a water circulation system, and before compression or testing, the samples were pre-dried in a well-ventilated oven at 40$ ^\circ$C for 4 hours.

\FloatBarrier
\section{Numerical methods}
Flow dynamics in material extrusion problems are complex due to the combination of shear and extensional flows, and these contributions cannot be decoupled experimentally, necessitating computational tools like Computational Fluid Dynamics (CFD). For numerical simulations in this project, OpenFOAM\R\, is employed due to its versatility, transparency, and open-source nature, allowing precise parameter manipulation \cite{openfoam}. The availability of solvers for viscoelastic fluids (rheoTool \cite{rheoTool}) in OpenFOAM\R\, allows getting more realistic results.

This section will introduce the numerical approach, from modelling the fluid domain and the polymer melts to the simulation of the viscoelastic flow in the extrusion nozzle. The numerical work is based on our previous published results \cite{schuller_add_ma} and it started with the 3D design and optimal meshing of the fluid domain, the definition of the constitutive models and, subsequently, the search for parameters that fit best the working fluids.

\subsection{Geometry modelling}

\begin{figure}[ht!]
\centering
    \includegraphics[width=0.4\textwidth]{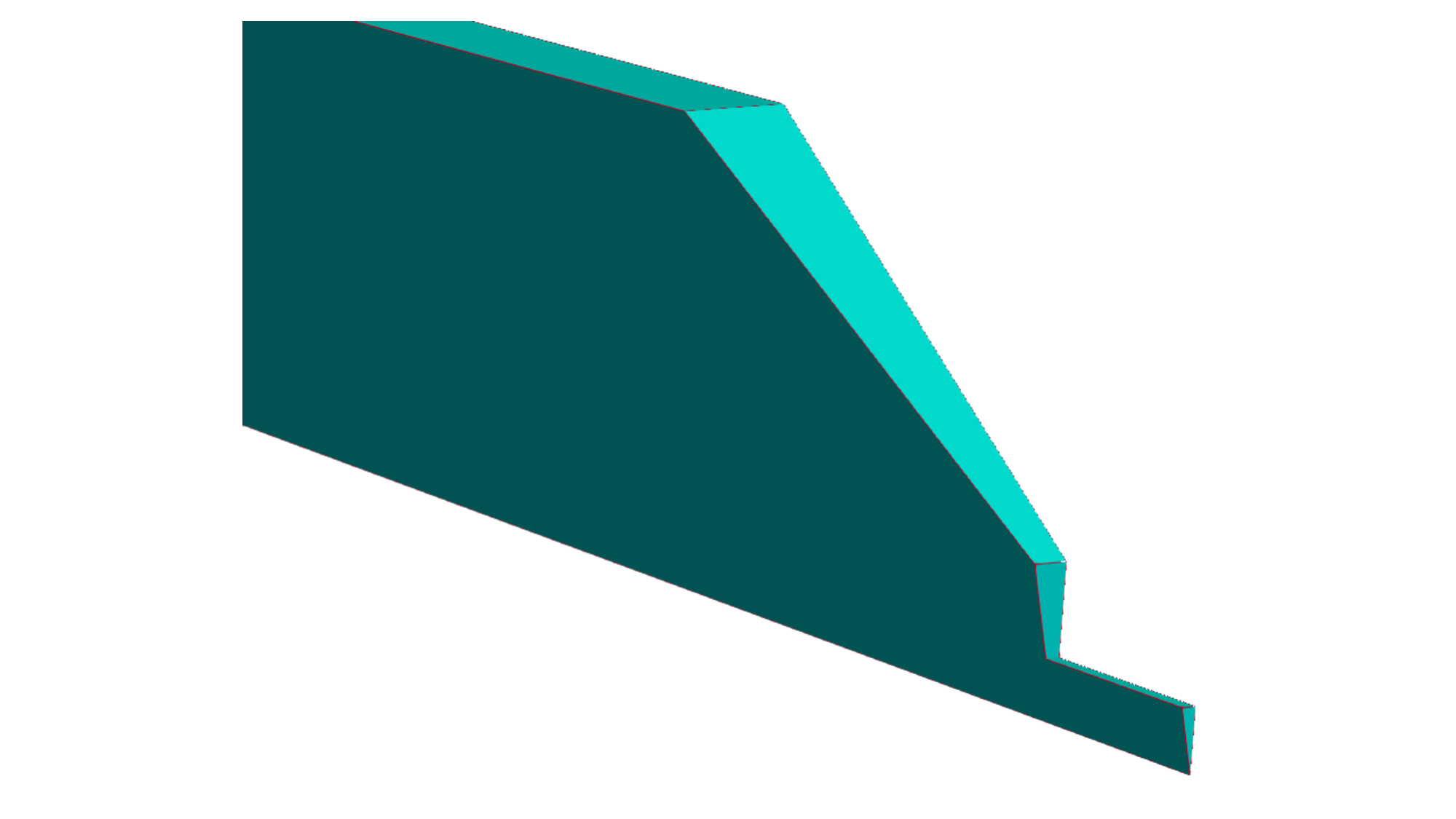}
    \caption{Isometric representation of the axisymmetric nozzle (AA).}
    \label{fig:7}
\end{figure}

Due to its conical shape, an axisymmetric geometry was deemed suitable for the nozzle. This decision resulted in geometry and mesh challenges, as the default cell structure of OpenFOAM\R\, is hexahedral. 
Due to the geometry's inherent axisymmetry, we seized an opportunity to realize substantial computational time reductions by choosing two-dimensional numerical simulations over 3D ones. Consequently, we set up the geometry as a thin wedge with a small angle and a single-cell width along the central axis. We designated distinct wedge-type patches for velocity and pressure as the boundary conditions on the axisymmetric wedge planes to accommodate this axisymmetric configuration.
To create an axisymmetric section (wedge), the external utility \textit{wedgePlease} \cite{wedgePlease} was utilized, resulting in the generation of a 5-degree angle section of the simplified nozzle assembly. This geometry can be observed in Figure \ref{fig:7}.

\subsection{Constitutive model}

The constitutive model needs to function efficiently in non-linear viscoelastic regimes to obtain reliable numerical results of non-Newtonian fluid flows. This model must consider shear-thinning and normal stresses to provide precise predictions of rheometric data.
The Giesekus model, as outlined in references \cite{GIESEKUS198269,GIESEKUS1985349}, presents a non-linear constitutive framework described by the subsequent equation:

    \begin{equation}
        \boldsymbol{\tau} + \lambda \overset{\triangledown}{\boldsymbol{\tau}}+\alpha\frac{\lambda}{\eta_0}\left(\boldsymbol{\tau}\cdot\boldsymbol{\tau}\right) =\eta_0 \left(\nabla\boldsymbol{u}+\nabla\boldsymbol{u}^T\right),       \label{Eq:giesekus}
    \end{equation}

\noindent where $\boldsymbol{\tau}$ represents the extra stress tensor and $\overset{\triangledown}{\boldsymbol{\tau}}$ its upper convected derivative, $\lambda$ stands for the relaxation time, $\alpha$ serves as the dimensionless mobility factor, $\eta_0$ signifies the zero-shear viscosity, and $\boldsymbol{u}$ denotes the velocity field in the Giesekus model. $\alpha$ plays a crucial role in governing both the extensional viscosity and the ratio of the second normal stress difference to the first.

\noindent The Giesekus model predicts shear-thinning, as well as it does predict shear 1\textsuperscript{st} and 2\textsuperscript{nd} normal-stress differences; however it predicts overshoots in shear and 1\textsuperscript{st} normal stresses in start-up experiments, which are a kind of transient experiments that are not involved in the cases analysed here \cite{understanding, larson1988, Maja}.

To determine model parameters, \textit{rheoTestFoam} was used to compare numerical responses set under the same experimental conditions with the rheometric data.

\subsection{Nozzle flow simulations}

The nozzle flow simulations were configured following the guidelines outlined in \cite{schuller_add_ma}. The specified boundary conditions are as follows: a non-slip condition and zero pressure gradient were enforced at the inner wall of the nozzle. Additionally, a uniform velocity profile ($\boldsymbol{V_{in}}=\overline{V_{in}}\boldsymbol{e_z}$) was prescribed at the inlet surface, where the average velocity $\overline{V_{in}}$ is linked to the extrusion velocity $\overline{V_{ext}}$ through the following relationship:

\begin{equation}
    \overline{V_{in}}= \frac{Q}{A_{fil}} =\overline{V_{ext}}\frac{D_c^2}{D_u^2}.
\end{equation}

\noindent where $D_c$ is the contraction diameter and $D_u$ is the liquefier diameter.

At the inlet, a zero-pressure gradient was imposed, and at the die's exit, where the extruded material exits to the atmosphere, the pressure and velocity boundary conditions were set to zero gradients in accordance with the approach outlined in a previous study by Schuller et al. \cite{schuller_add_ma}.

For mesh generation, the \textit{blockMesh} utility was utilized, creating three zones: the straight upstream region, the tapered region, and the die region.
To enhance the resolution at the die and the abrupt contraction areas, a streamwise stretching ratio was applied, keeping the same mesh size as in previous studies \cite{schuller_add_ma}.

Isothermal and steady flow through the extrusion nozzle was assumed for all materials and simulations. Consequently, the energy equation was decoupled from the mass and momentum conservation equations. Treating the molten polymer as an incompressible fluid, the mass conservation equation was simplified as depicted in Equation \ref{eq: Continuity}:
 
 \begin{equation}
  \label{eq: Continuity}
  \nabla \cdot \boldsymbol{u} =0. 
\end{equation} 

\noindent Equation \ref{eq: Momentum} gives the momentum conservation equation:
 
\begin{equation}
  \label{eq: Momentum}
  \rho \boldsymbol{u}\cdot\nabla \boldsymbol{u} =-\nabla P -\nabla\cdot\boldsymbol{\tau}.
\end{equation} 

\noindent In this context, we considered a steady-state flow ($\frac{\partial\boldsymbol{u}}{\partial t}=0$), while excluding the negligible effects of gravity. Here, we used the notation $P$ for pressure, $\rho$ for density, and the extra stress tensor $\boldsymbol{\tau}$, which encompasses contributions arising from the fluid deformation as defined in the Giesekus model found in Eq. \ref{Eq:giesekus}. To solve Eqs. \ref{eq: Continuity}, \ref{eq: Momentum}, and \ref{Eq:giesekus}, we utilized the rheoTool library~\cite{rheoTool,pimenta1} for OpenFOAM\R. In order to ensure numerical stability when dealing with the viscoelastic fluid, both the log-conformation formulation of the constitutive equation and a high-resolution scheme (CUBISTA) were employed~\cite{AlvesetalAnnRevFluMech2021}.

\FloatBarrier
\section{Results and Discussion}

\begin{figure}[ht!]
\centering
    \includegraphics[width=0.85\textwidth]{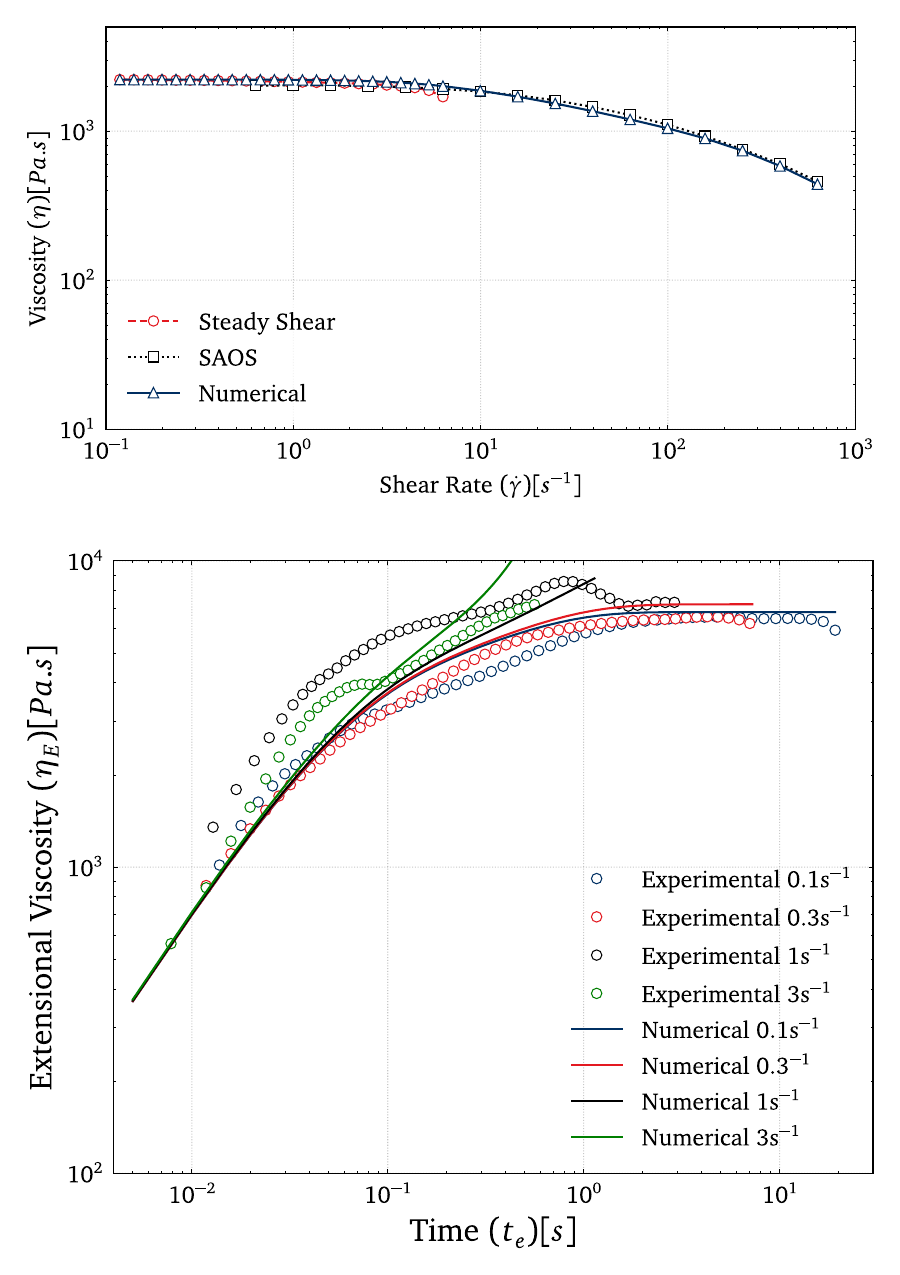}
    \caption{Experimental \textit{vs} numerical viscosity curves for PLA – shear and extensional.}
    \label{fig:visc_pla}
\end{figure}

The results of the rheological tests are depicted in Figure~\ref{fig:visc_pla}. Notably, it is possible to observe, while at low shear-rates, a zero-shear viscosity plateau, followed by a region of shear thinning above approximately 10 s$^{-1}$. Moreover, the dynamic and steady shear measurements were coupled by using the Cox-Merz rule \cite{CoxMerzrule}.\\

It is worth noting that lower extension rates could not be achieved due to the sagging of the samples during the test. The samples were preheated for 30 seconds at the prescribed temperature, which was followed by applying an 0.8 mm pre-stretch at a rate of 0.005 s$^{-1}$, and then maintaining a relaxation time of 5 seconds. Figure~\ref{fig:visc_pla} illustrates the transient evolution of extensional viscosity ($\eta^{+}_{E}\left(t,\dot\epsilon\right)$ \cite{nomenclatureJOR,rev1,rev2}) at the pre-determined extension rates. Remarkably, a linear behavior is observed, with overlap between the four curves. No significant extensional strain hardening \cite{Tervoort,Govaert,LIN2016143} is evident.

Unfortunately, due to the limitations of this experimental arrangement, it was not possible to determine the steady extensional viscosity at every extension rate. As a result, we cannot conclusively determine whether the material demonstrates extension thickening (an increase in $\eta_E$ with $\dot\epsilon$) or extension thinning (a decrease in $\eta_E$ with $\dot\epsilon$) as described in prior research~\cite{DEALY2018369}.

Table \ref{tab:prop_pla} details the values of the parameters of the polymer model, and Figure \ref{fig:visc_pla} shows the experimental shear and extensional viscosity plots compared with the numerical predictions.

\begin{table}[ht!]
    \centering
    \caption{Polymer melt parameters - PLA.}
    \begin{tabular}{ccccccc} 
    \textbf{Material}  & \textbf{Model} & \textbf{Num. Modes} & \textbf{Mode} & $\boldsymbol{\eta_0}$& $\boldsymbol{\lambda}$ & $\boldsymbol{\alpha}$\\ \hline
    \multirow{3}{*}{PLA} & \multirow{3}{*}{Giesekus} & \multirow{3}{*}{3} & 1 & 750 & 0.04 & 0.0015   \\  
     & & & 2 & 500 & 0.09 & 0.018 \\
     & & & 3 & 970 & 0.4 &  0.02
    \end{tabular}
    \label{tab:prop_pla}
\end{table}

\begin{figure}[ht!]
\centering
    \includegraphics[width=0.9\textwidth]{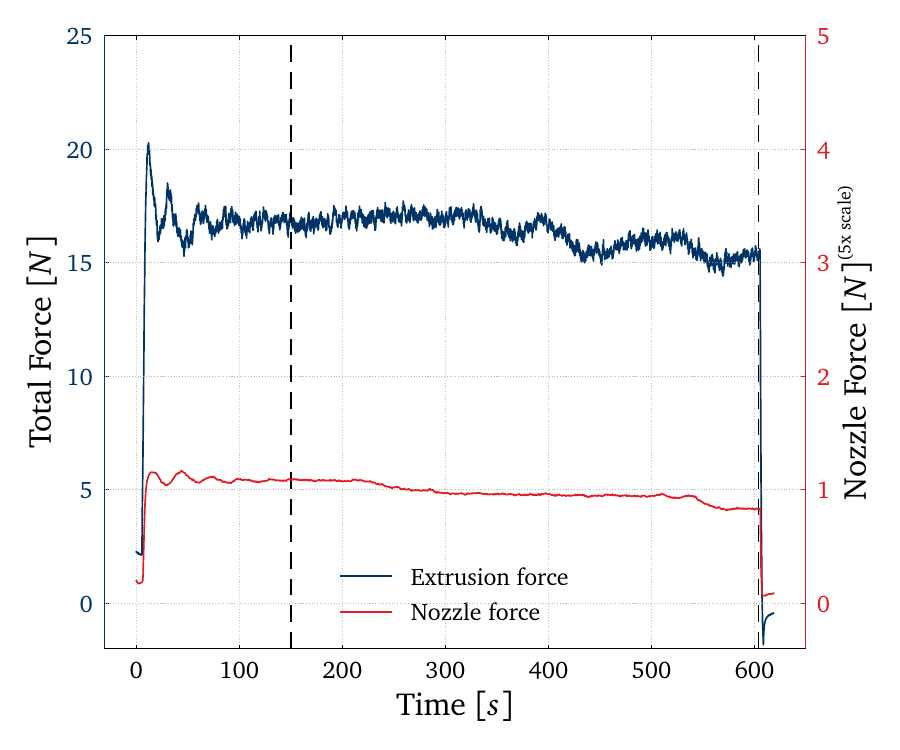}
    \caption{Typical data recorded at a flow of 1 mm$^3$/s and temperature of $205\,^\circ $C. The plot shows the recorded force signal from the load cell positioned between the feeder and the print-head $F_{tot}$ (blue, left y-axis), and the load cell connected with the sensorized nozzle $F_n$ (red, right y-axis). The average force is calculated between the dashed lines.}
    \label{fig:trace}
\end{figure}

Extrusion experiments have been performed using the sensorized nozzle. Figure~\ref{fig:trace} shows a typical recording of the force signal (blue line) acquired with the load cell positioned between the feeder and the print-head ($F_{tot}$) and the force signal (red line) from the force sensor connected to the pin in the nozzle ($F_n$). As expected, it can be observed that the measured total extrusion force is considerably larger than the nozzle force. 
Force measurements were stable and reproducible, showing similar results after testing at least three different nozzles having the same inner geometry and using the same extrusion conditions (volumetric flow rate and temperature values). Some leakage has been experienced during extrusion, but its effect has been considered negligible (a complete analysis of the accuracy of the force measurements can be found in Supplementary Information).\\

\begin{figure}[ht!]
\centering
    \includegraphics[width=\textwidth]{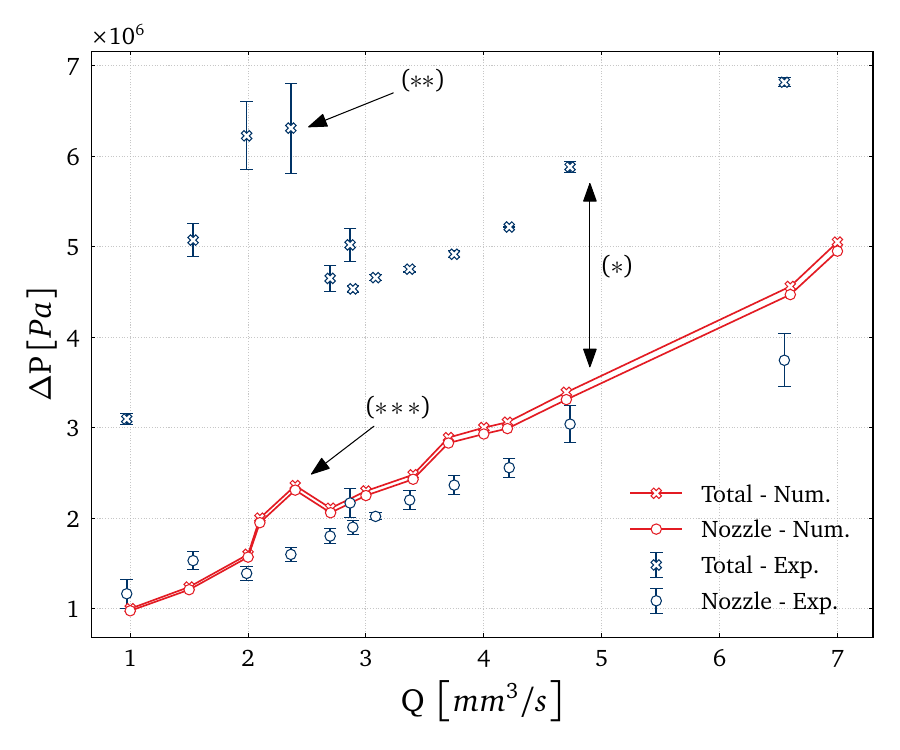}
    \caption{Average BB nozzle pressure drop (dots) and total extrusion pressure drop (X symbols) calculated from the experimental data (blue) and numerical pressure drop (red). Error bars represent standard deviation.}
    \label{fig:10}
\end{figure}

From the force measurements, the total pressure drop ($\Delta $P$_{tot}$) and nozzle pressure drop ($\Delta $P$_n$), as shown in Figure~\ref{fig:2}, were calculated using the equations $\Delta $P$_n = F_n / A_{pin}$, where $A_{pin}$ is the frontal area of the pin and $\Delta $P$_{tot} = F_{tot} / A_{fil}$, where $A_{fil}$ is the cross-section area of the solid polymer entering in the feeder. \\
Figure~\ref{fig:10} shows the pressure drop in the nozzle $\Delta $P$_n$ (dots) and the total pressure drop $\Delta $P$_{tot}$ (X symbols) as a function of the input volumetric flow rate at 205 \textdegree C. Experimental data (blue) is plotted against the numerical data (red).
As expected, the pressure drop in the nozzle is lower than the total pressure drop for all the flow rates tested since the latter accounts for the pressure drop in the nozzle and in the liquefier, $\Delta $P$_{liq}$, (see Figure~\ref{fig:2}): 
\begin{equation}
\Delta \textrm{P}_{tot} = \Delta \textrm{P}_{liq} + \Delta \textrm{P}_n.
\label{Eq:TotPres}
\end{equation}
However, some features in Figure~\ref{fig:10} are particularly striking:
\begin{enumerate}
    \item The measured total pressure drop ($\Delta $P$_{tot}$) is much higher than numerically predicted (*), and the nozzle pressure drop is not the biggest contributor to the total pressure drop (Eq.~\ref{Eq:TotPres}). Conversely, the numerical simulations match the experimental results quite well for the nozzle pressure drop. This means that the numerical model accurately describes the physics of the flow within the nozzle, where the model's assumptions (polymer fully melted and constant temperature) are satisfied. On the contrary, the model does not predict the pressure drop in the liquefier. Moreover, the total extrusion pressure drop exhibits non-linear dependence with the imposed flow rate, with a peak at 2 mm$^3$/s (**), which is not present in the simulation.
    \item Both the simulated (red) and measured (blue) nozzle pressure drop, $\Delta $P$_{n}$, exhibit a non-linear dependence with the imposed flow rate. An unexpected increase appears respectively at 2.5 mm$^3$/s and 1.5 mm$^3$/s (***).\\
\end{enumerate}
In the next two sections, these two aspects are further discussed.

\FloatBarrier
\subsection{Evaluation of the physical phenomena in the liquefier: backflow}

In this section, we analyze the first features that have been observed in Figure~\ref{fig:10}: high total extrusion pressure drop compared to simulated one (*) and non linearities in the total pressure drop (**).
Since those effects cannot be seen in the nozzle pressure drop, we exclusively focus on the pressure drop in the liquefier ($\Delta $P$_{liq}= \Delta $P$_{tot} - \Delta $P$_n$). The numerical simulations predict values of $\Delta $P$_{liq} $ ranging from $2 \cdot 10^4$~Pa to $10^5$~Pa, while the experimental data are of one order of magnitude higher, in the range between $2 \cdot 10^6$~Pa and $5 \cdot 10^6$~Pa.
Knowing that the measurement of the nozzle pressure drop agrees well with the numerical simulations, it can be concluded that the mismatch in the liquefier pressure drop is due to the inability of the numerical model to capture two main physical aspects of the material extrusion printing process:
\begin{enumerate}
    \item The high-temperature gradient present in the liquefier since the solid polymeric filament enters the print-core at room temperature and is heated to $205\,^{\circ}$C, whereas the simulations assume a fully melted polymer entering the liquefier at a constant temperature ($205\,^{\circ}$C). 
    \item The backflow of polymer towards the annular region between the solid filament and the liquefier wall. In our previous study \cite{schuller_add_ma}, it was suggested that the dominating shear-induced normal stresses in that region could potentially be accountable for an additional pressure drop in the liquefier, which was not incorporated into the numerical model.
\end{enumerate}

Regarding the first aspect, the temperature gradient in the liquefier cannot explain such a big difference in $\Delta P_{liq}$ between the experimental and simulated results. Assuming that the pressure drop in a tube is proportional to the viscosity, the PLA real viscosity should be 50 times bigger than the one used for the simulations. Therefore, a reduction in the liquefier temperature cannot explain such a big increase in viscosity since it has been measured \cite{rheoT} that the viscosity at $180\,^{\circ}$C is around 4.5 times higher than the one at $210\,^{\circ}$C.

So, we can consider that the major contribution to the high measured values of the liquefier pressure drop is the backflow.\\

\begin{figure}[ht!]
     \centering
     \begin{subfigure}[t]{0.49\textwidth}
         \centering
         \includegraphics[width=\textwidth]{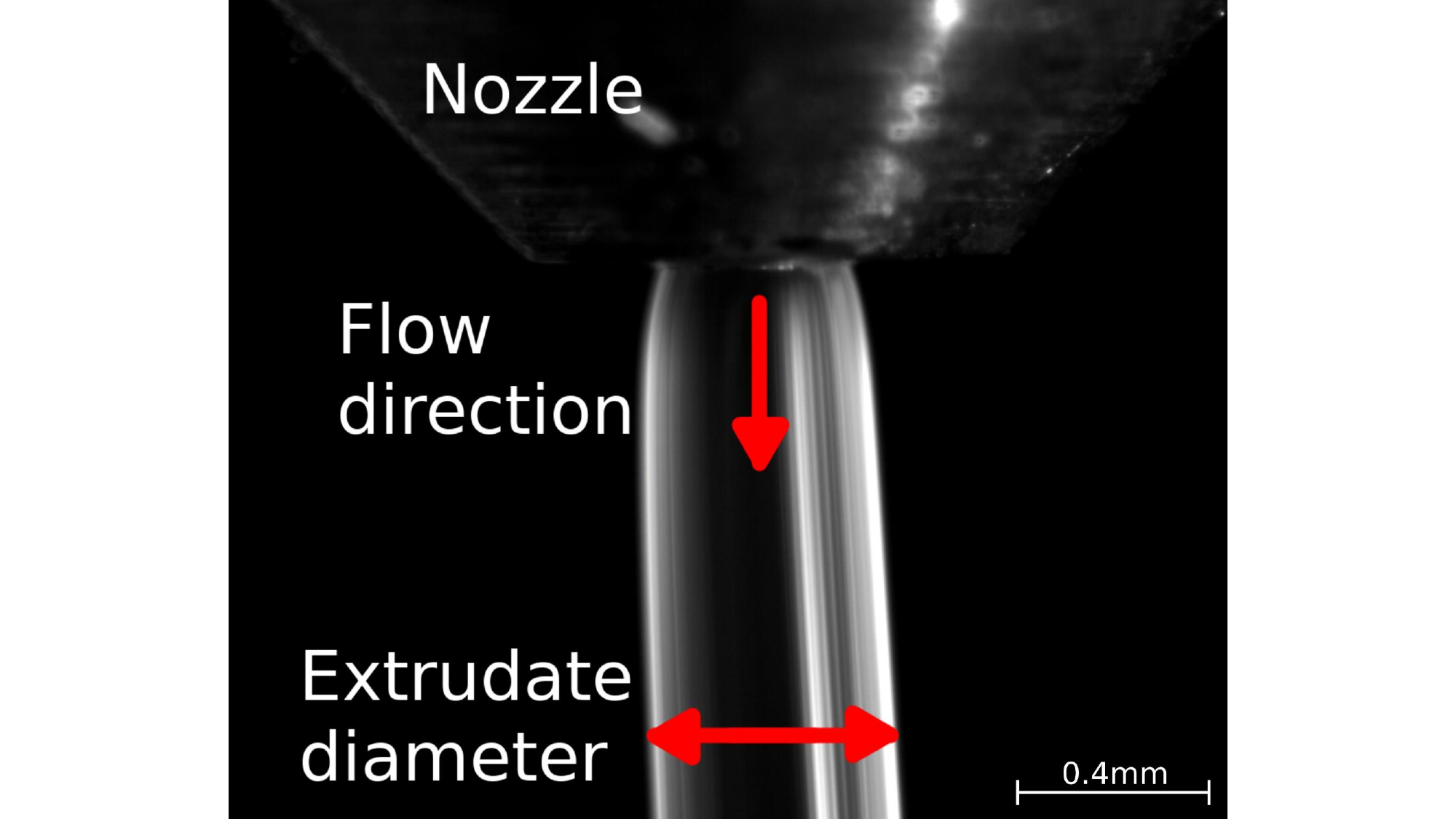}
         \caption{Example of a typical image of the extrudate exiting the BB nozzle.}
         \label{fig:dieswell_fig}
     \end{subfigure}
     \hfill
     \begin{subfigure}[t]{0.5\textwidth}
         \centering
         \includegraphics[width=\textwidth]{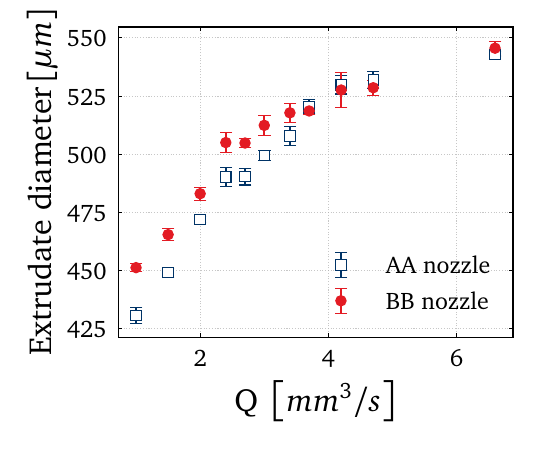}
         \caption{Average extrudate diameter for AA (blue) and BB (red) shape nozzles. Error bars represent standard deviation.}
         \label{fig:dieswell_plot}
     \end{subfigure}
        \caption{Die swelling measurement.}
        \label{fig:dieswell}
\end{figure}

Building on the findings from our prior research, we assumed that the variation between the observed total pressure ($\Delta $P$_{tot-Exp}$) and the forecasted value by the VE model ($\Delta $P$_{tot-VE}$) could be ascribed to the cumulative friction generated by the backflow ($\Delta $P$_{BF}$). The determination of the equilibrium height within the annular gap followed this approach \cite{schuller_add_ma}:
\begin{equation}
\label{eq:original}
 H^{\star}_{original} = \frac{\Delta \textrm{P}_{tot-Exp}-\Delta \textrm{P}_{tot-VE}}{(N_{1-BF_{numerical}} + \tau_{w-BF}) \cdot \frac{4}{D_f}}, 
\end{equation}
\noindent where $D_f$ is the filament diameter, $N_{1-BF}$ is the shear-induced normal stress difference, and $\tau_{w-BF}$ is the shear stress in the backflow region (measured at the wall of the filament).
$\tau_{w-BF}$ is calculated using the equation $\tau_{w-BF} = \eta \cdot \dot\gamma_{w-BF}$ where $\eta$ is the viscosity at the shear rate $\dot\gamma_{w-BF}$ determined analytically by Fakhari and Galindo-Rosales \cite{ahmad_curro_pof}:

\begin{equation}
\label{eq:bf_gammadot}
{\dot\gamma_{w-BF} = 2v(t) \frac{-8D_f^4ln(D_f/D_u)+D_u^4-6D_u^2D_f^2+5D_f^4}{D_f((D_u^4-D_f^4)ln(D_f/D_u)+(D_u^2-D_f^2)^2},}
\end{equation}

The dependence of the first normal stress difference $N_{1}$ with the shear rate is calculated from the die swelling experiments at different extrusion velocities (see Supplementary Information) using the R.I. Tanner~\cite{Tanner1970} equation:

\begin{equation}
    \frac{D}{D_c}=\left[1+\frac{1}{2}\left(\frac{N_{1}}{2\dot{\gamma}}\right)_{w}^{2}\right]^{1/6},
    \label{eq:tanner}
\end{equation}

\noindent \noindent where $\dot{\gamma}$ is the shear rate and the subscript $w$ prescribes that both $\dot{\gamma}$ and $N_{1}$ are determined at the wall exit of the die. Thus, $N_{1-BF}$, the shear-induced normal stress difference in the backflow region is then calculated from this dependence and using $\dot\gamma_{w-BF}$ from Eq.~\ref{eq:original}.

Figure~\ref{fig:dieswell} a) shows a typical optical image recorded during the extrusion of PLA. The extrudate diameter is evaluated at a distance of $500\,\mu$m from the nozzle tip and an average value is calculated using different frames.Figure~\ref{fig:dieswell} b) shows the average extrudate diameter in function of the flow rate for the BB (dots) and AA (squares) nozzle shape.

Thanks to the sensorized nozzle, it is now possible to measure experimentally the pressure drop in the nozzle ($\Delta $P$_{n-Exp}$), which is very close to the pressure drop predicted by the numerical simulations with the viscoelastic model ($\Delta $P$_{tot-VE}$). Thus, the value of $H^{\star}$ can be estimated just using the experimental data set as follows:

\begin{equation}
\label{eq:experimental}
 H^{\star}_{experimental} = \frac{\Delta \textrm{P}_{tot-Exp}-\Delta \textrm{P}_{n-Exp}}{(N_{1-BF_{Exp}} + \tau_{w-BF}) \cdot \frac{4}{D_f}}, 
\end{equation}

\begin{figure}[ht!]
\centering
\begin{subfigure}{.5\textwidth}
    \centering
    \includegraphics[width=\textwidth]{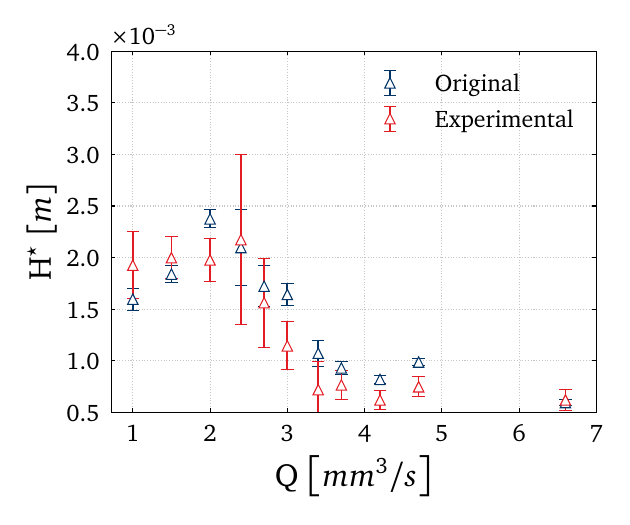}
    \caption[short]{}
\end{subfigure}%
\begin{subfigure}{.5\textwidth}
    \centering
    \includegraphics[width=\textwidth]{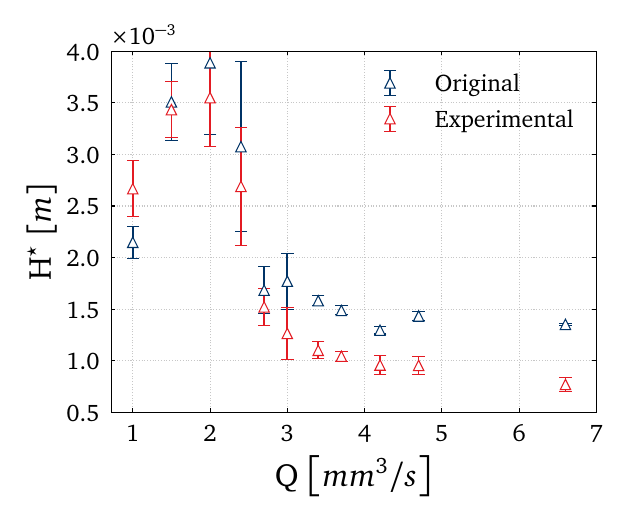}
    \caption[short]{}
\end{subfigure}
\caption[short]{Height of the meniscus $H^{\star}$ for AA (a) and BB (b) print-core.}
\label{fig:H}
\end{figure}

Figure~\ref{fig:H} shows a comparison between the backflow height calculated in these two ways for the AA (Figure~\ref{fig:H} a) and BB (Figure~\ref{fig:H} b) nozzle geometries. 
Since $ H^{\star}_{experimental}$ is calculated using only experimental data using the sensorized nozzle, we can expect it to better capture the the backflow phenomena. 
Since it can be observed that the values of $H^{\star}$ calculated using Eqs. \ref{eq:original} and \ref{eq:experimental} are very similar, we can confirm that equation \ref{eq:original} is valid and it can be used for the evaluation of the backflow phenomena. 
This is a remarkable result considering that the experimental data required to calculate $H^{\star}_{original}$ can be easily acquired. This means that Eq. \ref{eq:original} can be used for screening and characterizing the backflow behaviour for a broad range of polymers relevant to material extrusion.

\noindent As previously mentioned, an interesting aspect is the non-linearity of the liquefier pressure drop in function of the extrusion flow (Figure~\ref{fig:10} (**)).
Now, with the calculation of the backflow height, we can provide an explanation for such non-linearity: at low flows (below 2 mm$^3$/s), the backflow height increases with the flow (backflow is dominated by pressure-driven flow) while at high flows (above 2 mm$^3$/s), the backflow height decreases with the flow (backfow is dominated by drag flow of the solid filament entering in the liquefier). The balance between the two effects is reached at the peak, at around 2 mm$^{3}$/s . \\
Interestingly, the peak value of $H^{\star}$ is higher for the BB nozzle shape. According to Eq.~\ref{eq:experimental}, the value of $H^{\star}$ depends on the pressure drop in the nozzle ($\Delta $P$_{n-Exp}$). Therefore, modifications in the shape of the nozzle will affect the pressure drop and, consequently, the peak value of the $H^{\star}$ plot.

\begin{figure}[ht!]
\centering
\includegraphics[width=0.85\linewidth]{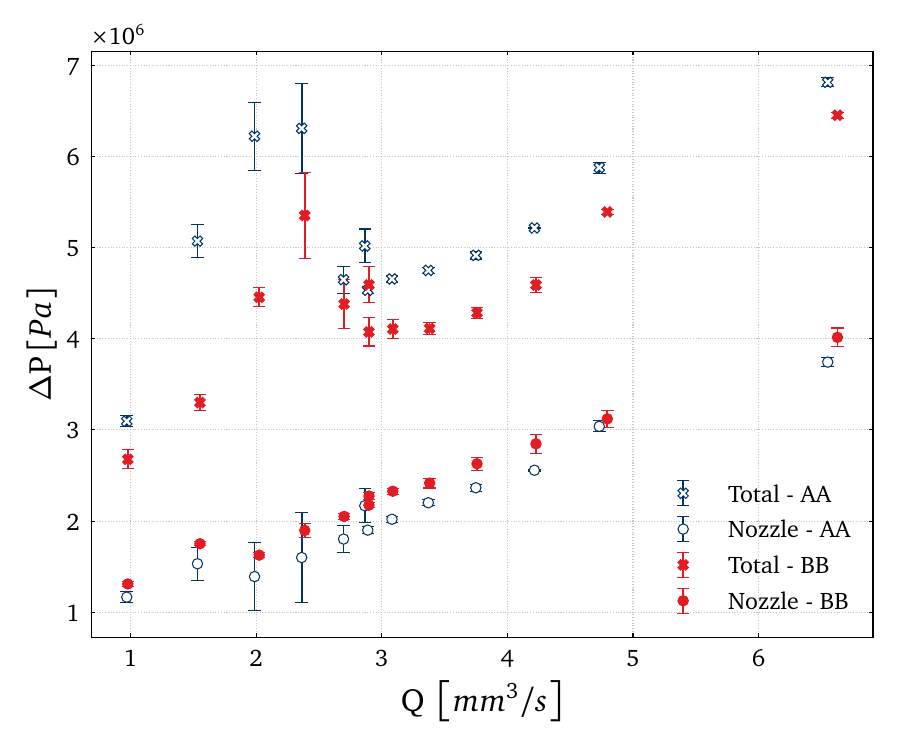}
\setlength{\belowcaptionskip}{-10pt}
\setlength{\abovecaptionskip}{-5pt}
\caption{Pressure drop at $205^{\circ}$C for AA (blue) and BB (red) shape nozzles. Error bars represent standard deviation.}
\label{fig:AABB}
\end{figure}

Figure~\ref{fig:AABB} compares the results obtained regarding pressure drop with the two different types of nozzle shapes. The peaks' presence and position are similar for both cases, further demonstrating that the non-linear effects are mainly in the annular gap of the liquefier while the nozzle shape has a minor influence on the height of the peak. 

\begin{figure}[ht!]
\centering
\includegraphics[width=0.7\linewidth]{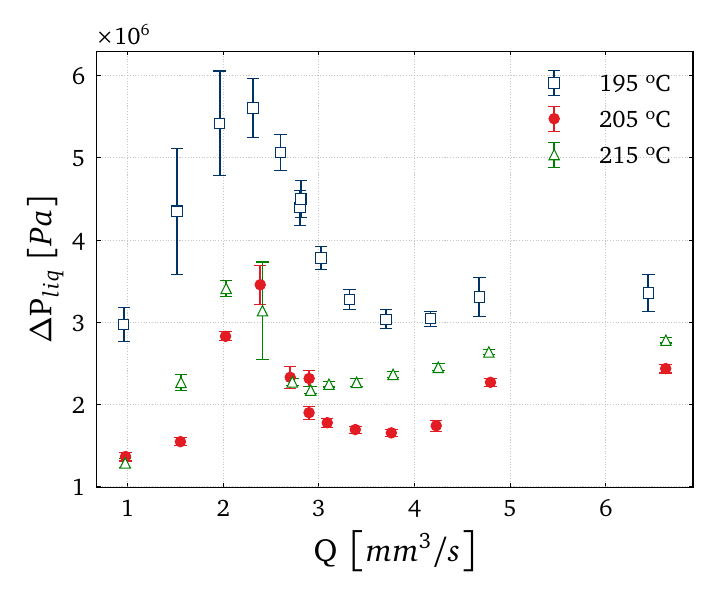}
\setlength{\belowcaptionskip}{-10pt}
\setlength{\abovecaptionskip}{-5pt}
\caption{Liquefier pressure drop at different extrusion temperature values for a BB nozzle shape. Error bars represent standard deviation.}
\label{fig:temp}
\end{figure}

\noindent This non-linear effect in the pressure drop data increases when decreasing the extrusion temperature as shown in Figure~\ref{fig:temp}, where the experimental liquefier pressure drop is plotted in function of the flow rate for $195\,^{\circ}$C, $205\,^{\circ}$C and $215\,^{\circ}$C. This is expected since decreasing the temperature results in an increase in the melt shear viscosity. Moreover, increasing the temperature allows for reducing the impact of the elastic behaviour of the polymer melt, as it can relax in a faster way. 
On one hand, an increment in the temperature will decrease the relaxation time and the viscosity of the molten polymer, therefore the equilibrium height should be shorter at 215 $^{\circ}$C than at 205 $^{\circ}$C and a lower pressure drop would be expected; however these effects were not observed in Figure \ref{fig:temp} \cite{Kang2019}. On the other hand, decreasing the temperature increases the viscosity and the relaxation time, and this results in a larger pressure drop sensed by the load cell located at the feeders (Figure \ref{fig:temp}), probably due to a higher equilibrium height .\\
That data shows the non-linear effect in the full range of printing temperatures. 

\noindent The same non-linear behaviour has also been observed in nozzles having the same shape but without the pin. \\
Those tests, performed with different nozzles and at different process conditions, demonstrate that the non-linearities in the extrusion pressure drop for PLA are present in the temperature range and flow range relevant in printing condition. Those effects cannot be neglected when optimizing the printing conditions and nozzle design. 

\FloatBarrier
\subsection{Non-linearities in nozzle pressure drop: elastic instabilities}

In this section we analyze the non-linearities observed in the nozzle pressure drop in Figure~\ref{fig:10} (***): in the simulation, a small peak appears at 2.5 mm$^3$/s.
During the simulations, the presence of instabilities originating from the re-entrant corner of the nozzle upon a critical extrusion velocity is observed. Consequently, these instabilities significantly impacted the accuracy and stability of pressure calculation. To address this issue, successive mesh refinements were undertaken, concluding that the numerical source was not responsible for the observed problem. In the experimental work of Rothstein and McKinley \cite{rothstein}, it was reported that there is a flow rate upon which elastic instabilities are triggered, yielding a range of values for pressure and recirculation dimensions upstream of the contraction section. When the Weissenberg number surpasses a critical value (Wi$_{crit}$ $\geq$ $\lambda_0 \langle \overline{V_z} \rangle_2 / R_2$, as per \cite{rothstein1}), steady numerical solutions are no longer achievable, and the flow transitions to a time-dependent state. These pure elastic instabilities, where inertia does not play a significant role, have also been observed experimentally and numerically in different geometries, including Taylor–Couette flow \cite{Larson1990,Schfer2018}, contraction flow \cite{McKinley1991,Alves2007}, and lid-driven cavity flow \cite{Pakdel1996,SOUSA2016129}, among others. It is widely acknowledged that the destabilizing mechanism responsible for these instabilities is a combination of substantial normal stresses (resulting in tension along the fluid streamlines) and the curvature of the streamlines. McKinley et al. \cite{McKinley1996} proposed a dimensionless parameter that must be exceeded for the onset of purely elastic instabilities derived from the combination of flow curvature and tensile stress along the streamlines. This criterion for the initiation of elastic instability, commonly known as the Pakdel-McKinley criterion, $\mathbb{M}$ \cite{Datta2022}, can be expressed in a general form as follows:

\begin{equation}
    \mathbb{M}=\sqrt{\frac{\lambda \mathbf{U}}{\mathcal{R}}\frac{\tau_{11}}{\eta \dot{\gamma}}}\,, 
\end{equation}

\noindent where $\lambda$ is the relaxation time of the fluid, $\mathbf{U}$ is the characteristic streamwise fluid velocity, $\mathcal{R}$ is the characteristic radius of curvature of the streamline, where $\tau_{11}$ is the primary normal stress component along the streamlines, $\dot{\gamma}$ is the shear rate and $\eta$ is the viscosity, which for simplicity was assumed as the zero-shear viscosity, that is the sum of each mode's viscosity from the multi-mode model.
For every fluid flow configuration, the value of the $\mathbb{M}$ parameter must remain below the critical value, which indicates when the elastic instabilities arise ($\mathbb{M}_{crit}$).

As depicted in \cite{Datta2022}, an extensive array of flows has been substantiated to exhibit elastic instability, primarily through experimental observations and measurements. Many of these flows involve curved streamlines, leading to instabilities attributed to hoop-stress effects. Consequently, they are typically characterized, to some extent, by the $\mathbb{M}$ parameter. The application of geometric scaling has yielded remarkable success. For instance, investigations have revealed that instability occurring in a serpentine channel flow is directly linked to the Dean instability \cite{Zilz2012}. Notably, the predicted dependence of the instability threshold in Taylor-Couette flow by the Pakdel-McKinley criterion better aligns with experimental values compared to the results obtained from linear stability analysis meticulously tailored to the fluid's rheology \cite{Schfer2018}.

\begin{figure}[ht!]
\centering
\includegraphics[width=0.9\linewidth]{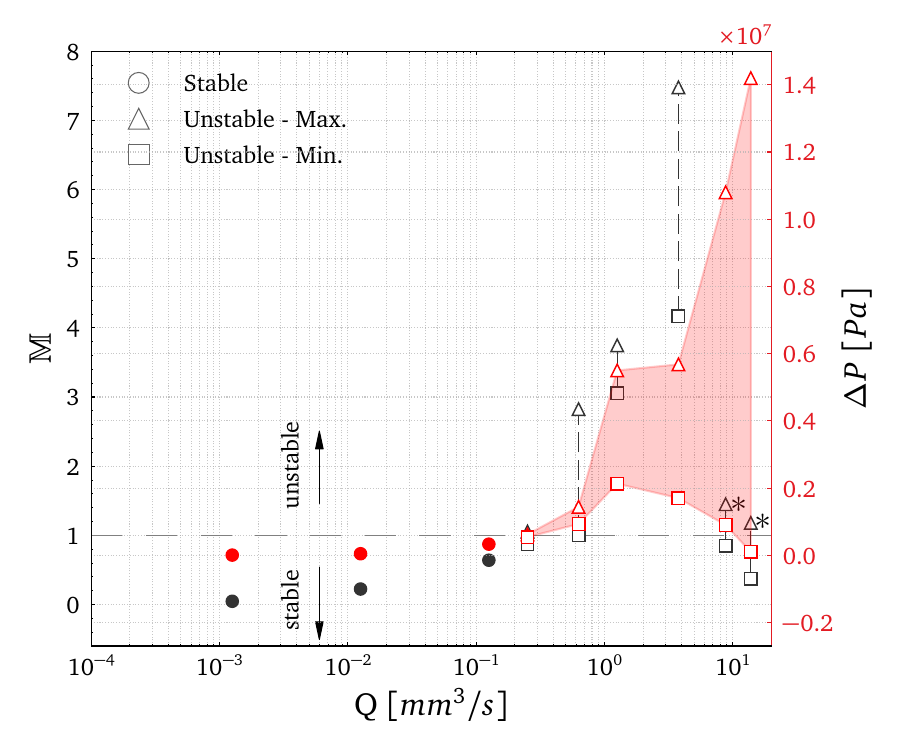}
 \caption[$\mathbb{M}$ (black) and total pressure drop (red) values \textit{vs} extrusion flow]{$\mathbb{M}$ (black) and total pressure drop (red) values \textit{vs} extrusion flow (numerically evaluated for the AA nozzle).\\{\Large\textsuperscript{$*$}}In the final two cases, $\mathcal{R}$ starts growing at a much larger rate compared to the streamwise velocity, leading to a muffling of the $\mathbb{M}$ value.}
\label{fig:mcrit_twinx}
\end{figure}

To calculate the curvature radius and the primary stresses in the streamline's coordinate system, a Python script has been developed to generate, with the input of the streamline point coordinates, a polynomial fitting and closest circumference to the desired point interval and to create a new coordinate system with the normal and tangent curvature vectors, that allow for the transformation of the cartesian tensions tensor to the new system \cite{foamscripts}. A graphical abridged version of this process is shown in Supplementary Information. Moreover, by using this script, it is possible to determine $\mathbb{M}$ values for the different extrusion velocities. Figure \ref{fig:mcrit_twinx} shows a twin-axis plot, where the Pakdel-McKinley criterion and total pressure values are plotted against the flow for PLA. It can be observed that the numerical simulations predict elastic instabilities in the tapered region of the BB core upon $\overline{V_{ext}}>1$~mm/s (Q $\approx$ 0.12 mm$^{3}$/s).

If we compare this result with the pressure drop graph in Figure~\ref{fig:10} (***), we can conclude that the peak at 2.5 mm$^3$/s is linked with the flow instability, whose effect increases starting from low flow rates and reaches an onset between 1 mm$^3$/s and 3 mm$^3$/s in the tapered region.
In the experimental data, the peak is shifted at 1.5 mm$^3$/s. Such a shift can be due to the not perfect fitting of the constitutive model to the rheometric data or the effect of temperature.

The above discussion explains the non-linearity in the simulated and measured nozzle pressure drop data ($\Delta $P$_{n}$). 
Nevertheless, according to the numerical prediction, it can be inferred that the non-linearities in the total pressure drop are present in the backflow region, where the shear-induced normal stresses are dominant. Although this idea is supported by the fact that shear-induced normal stresses can be measured in a small-scale slit-die~\cite{Teixeira2013661,Hilliou202124}, further research is needed to prove that they are dominant in the backflow region. Moreover, according to \cite{Machado2016}, elastic instabilities also occur in straight channels, resulting in a significant increase of the pressure drop above a critical Weissenberg value that might be interpreted as due to another type of elastic instability. Therefore, the onset of those elastic instabilities coincides with the measurement of an extra-pressure drop in the backflow region.
In conclusion, there are elastic instabilities both in the tapered region, that is something we can observe and calculate with the $\mathbb{M}$ parameter. Moreover, the numerical model predicts a slight overshoot in the pressures at the nozzle location, which is due to that elastic instability at the nozzle contraction region. This was not detected experimentally by the pin. However, we conclude that the discrepancy between the numerical and experimental total pressure drop is due to the backflow region. If it was just a matter of viscosity, then the rise in the pressure would be uniform; however we observe an increase in the nonlinearity as well. We speculated that the large overshoot measured in the total pressure drop is also due to elastic instabilities occurring at the backflow region.

\FloatBarrier
\section{Conclusions and future works}

In this study, a novel nozzle pressure monitoring method was developed and tested, proving that it can repeatedly measure the force inside the nozzle of a material extrusion printer. It consistently measured melt pressure across various nozzle designs, temperatures, and flow rates. Additionally, the obtained experimental results aligned well with numerical pressure calculations.
The proposed method for monitoring nozzle pressure in material extrusion represents a significant advancement in utilizing experimental pressure data to comprehend the intricate dynamics of extrusion processes, opening new avenues for enhancing extrusion control strategies. 
In particular, measuring the pressure drop at different locations along the extrusion path enables the discrimination and description of the dominant physical effects in the various sections of the extrusion print-head: the liquefier and the nozzle. 
This study successfully captured previously unobserved non-linearities in the pressure drop. Thanks to the nozzle pressure drop measurements, it was possible to conclude that the shear-induced elastic instabilities occurring in the backflow region of the liquefier are the source of the non-linearities captured by the pressure sensor located in the feeders. Moreover, the experimental determination of the nozzle pressure drop also allows for the calculation of equilibrium height in the backflow region based on experimental data exclusively, thus validating our previous results obtained using the numerical simulations.
This study purely focuses on the extrusion process, but it can lead to some recommendations for the development of print-head with improved performances and the optimization of the printing process. 
First of all, developing and validating a numerical model that matches the experimental data well poses the basis for the topological optimization of the nozzle design to minimize the elastic instabilities and the pressure drop. 
Additionally, the results of this paper underline the importance and the big contribution to the total pressure drop of the liquefier. Minimizing such extra-pressure drop due to the backflow is of fundamental importance when printing at high speed. In such conditions, the feeder must feed the filament at high speed, overcoming a high-pressure drop. This leads to a high feeder slip, which can be detrimental to the success of the printing process. So, minimizing the liquefier pressure drop is imperative to enable high printing speed. 
Another important implication of this work is that it enables to have a reliable model to identify the printing process window by looking at the flows and temperature values at which the non-linear phenomena are triggered. In particular, it is quite interesting to note that, for PLA, a low extrusion flow can be problematic due to the fact the backflow height has a maximum value at which the polymer melt could enter the cold zone, completely clogging the print-head. 
This also has implications in the selection of filaments for material extrusion applications, with rheological properties that do not show (or minimize) the presence of elastic instabilities, moving the non-linear peak of pressure drop at higher flows.
In conclusion, this paper represents a step forward in the path towards building further and unraveling the mechanics of the extrusion process in fused filament fabrication.

\section{CRediT authorship contribution statement}

\textbf{Sietse de Vries}: writing - review $\&$ editing, methodology, investigation, visualization, validation. 
\textbf{Tom\'as Schuller}: writing - review $\&$ editing, formal analysis, software, investigation, visualization, validation. 
\textbf{Francisco J. Galindo-Rosales}: writing - review $\&$ editing, funding acquisition, supervision, resources.
\textbf{Paola Fanzio}: writing - original draft, supervision, conceptualization, project administration, resources.

\section{Declaration of competing interest}

The authors declare that they have no known competing financial interests or personal relationships that could have appeared to influence the work reported in this paper.

\section*{Acknowledgements}
The authors would like to thank Francisco Vide Coelho de Almeida, Johan Versteegh, Koos Welling, Sam Kent and Can Ayas for the fruitful discussion and selfless support. FJGR and TS acknowledge the financial support from Ultimaker B.V., and also LA/P/0045/2020 (ALiCE), UIDP/00532/2020 (CEFT) and the program Stimulus of Scientific Employment, Individual Support-2020.03203.CEECIND, funded by national funds through FCT/MCTES (PIDDAC).

\FloatBarrier
\newpage

\section*{Supplementary Information for: "Pressure drop non-linearities in material extrusion additive manufacturing: a novel approach for pressure monitoring and numerical modeling"}

Sietse de Vries, Tom\'as Schuller, Francisco J. Galindo-Rosales, and Paola Fanzio.

\section[\appendixname~\thesection]{Characterization of the nozzle}
\label{app:char}
In this section, same examples of optical images of the nozzle components are presented. 
All the tested nozzles have been optically characterized. 
\begin{figure}[ht!]
\centering
\includegraphics[width=0.6\linewidth]{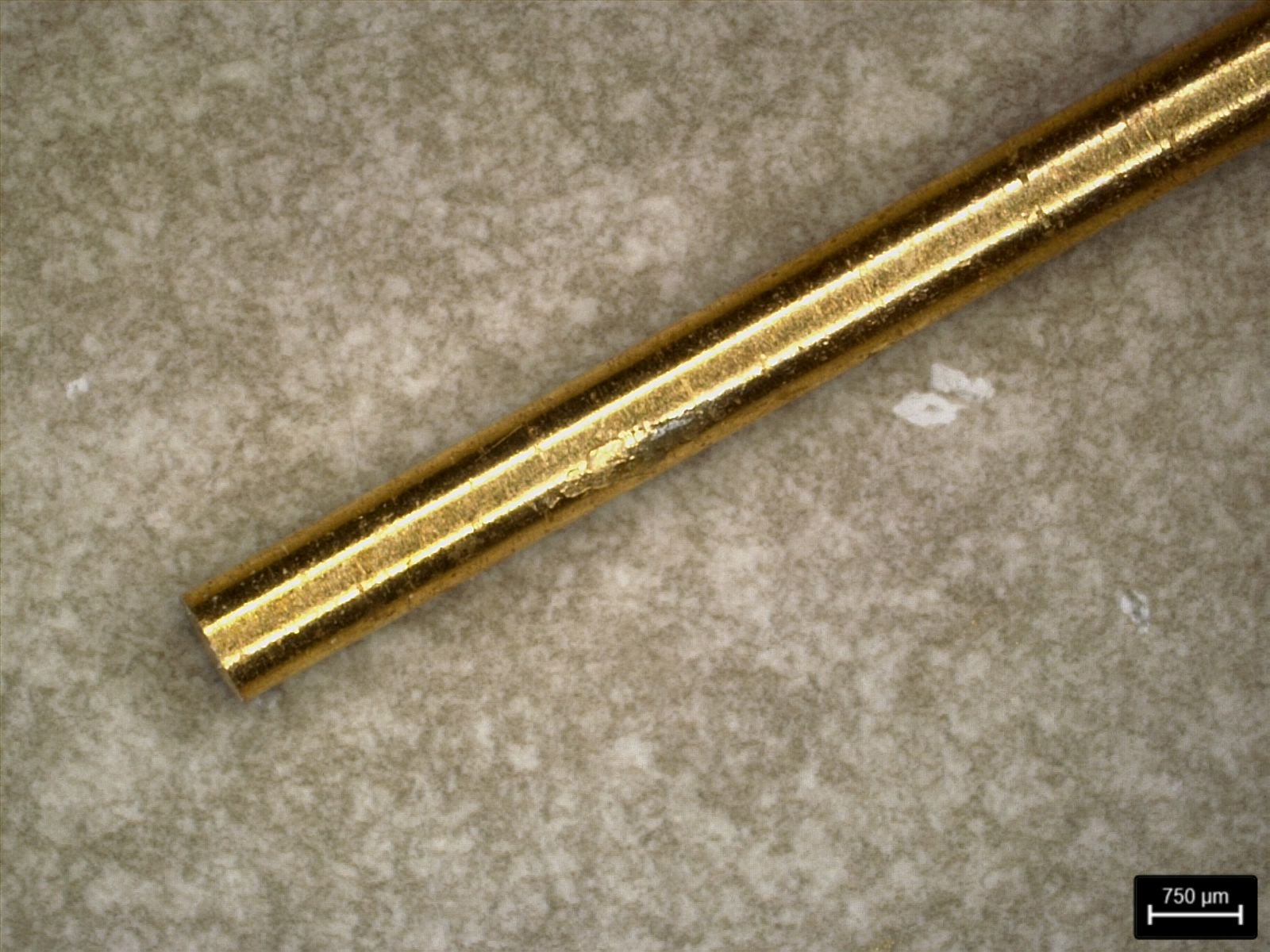}
\caption{Optical image of a typical pin.}
\label{fig:opt1}
\end{figure}

\begin{figure}[ht!]
\centering
\includegraphics[width=0.6\linewidth]{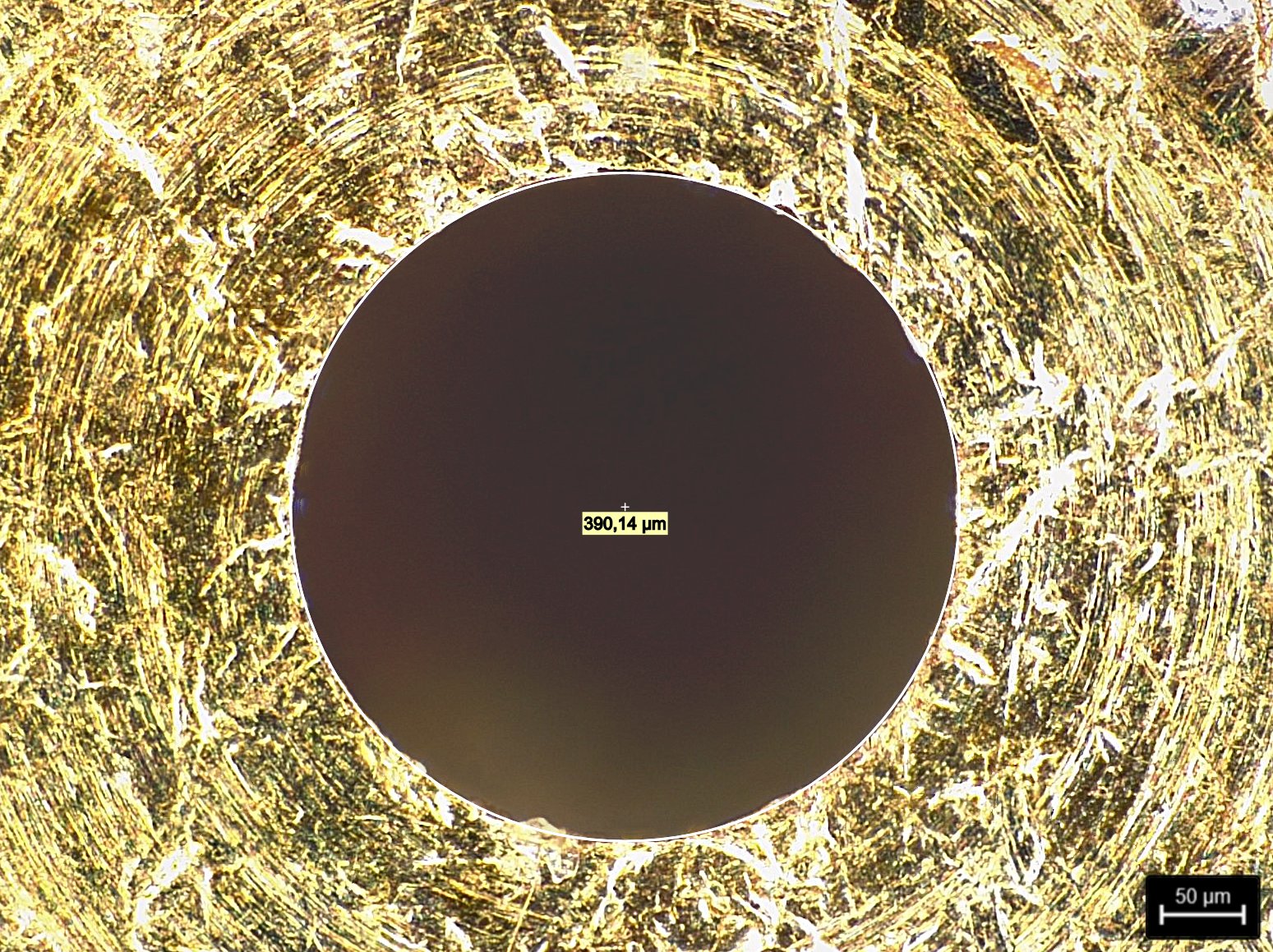}
\caption{Typical dimension of the nozzle die.}
\label{fig:opt2}
\end{figure}

\begin{figure}[ht!]
\centering
\includegraphics[width=0.5\linewidth]{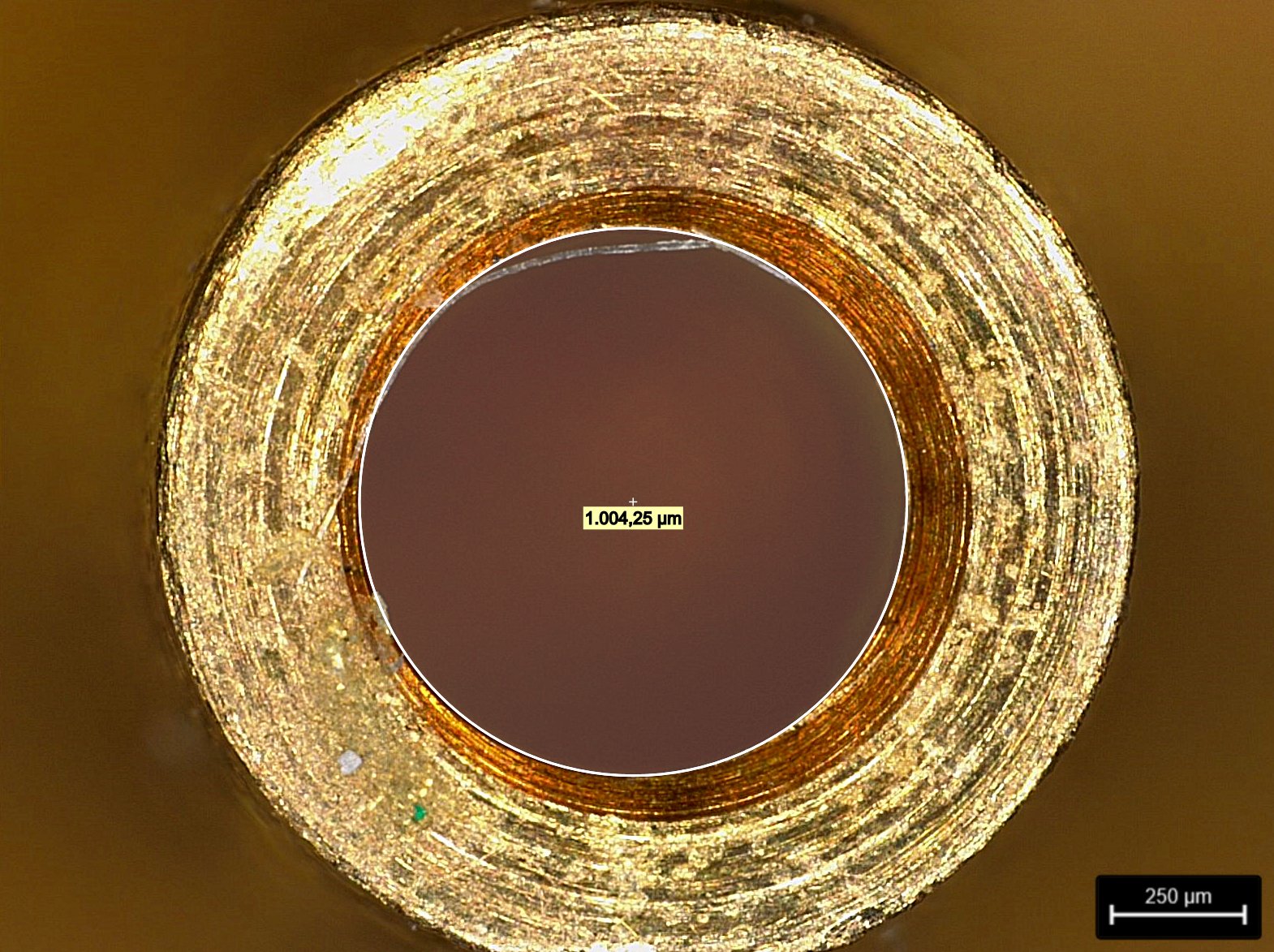}
\caption{Aperture in the guidance tube.}
\label{fig:opt3}
\end{figure}

\begin{figure}[ht!]
\centering
\includegraphics[width=0.5\linewidth]{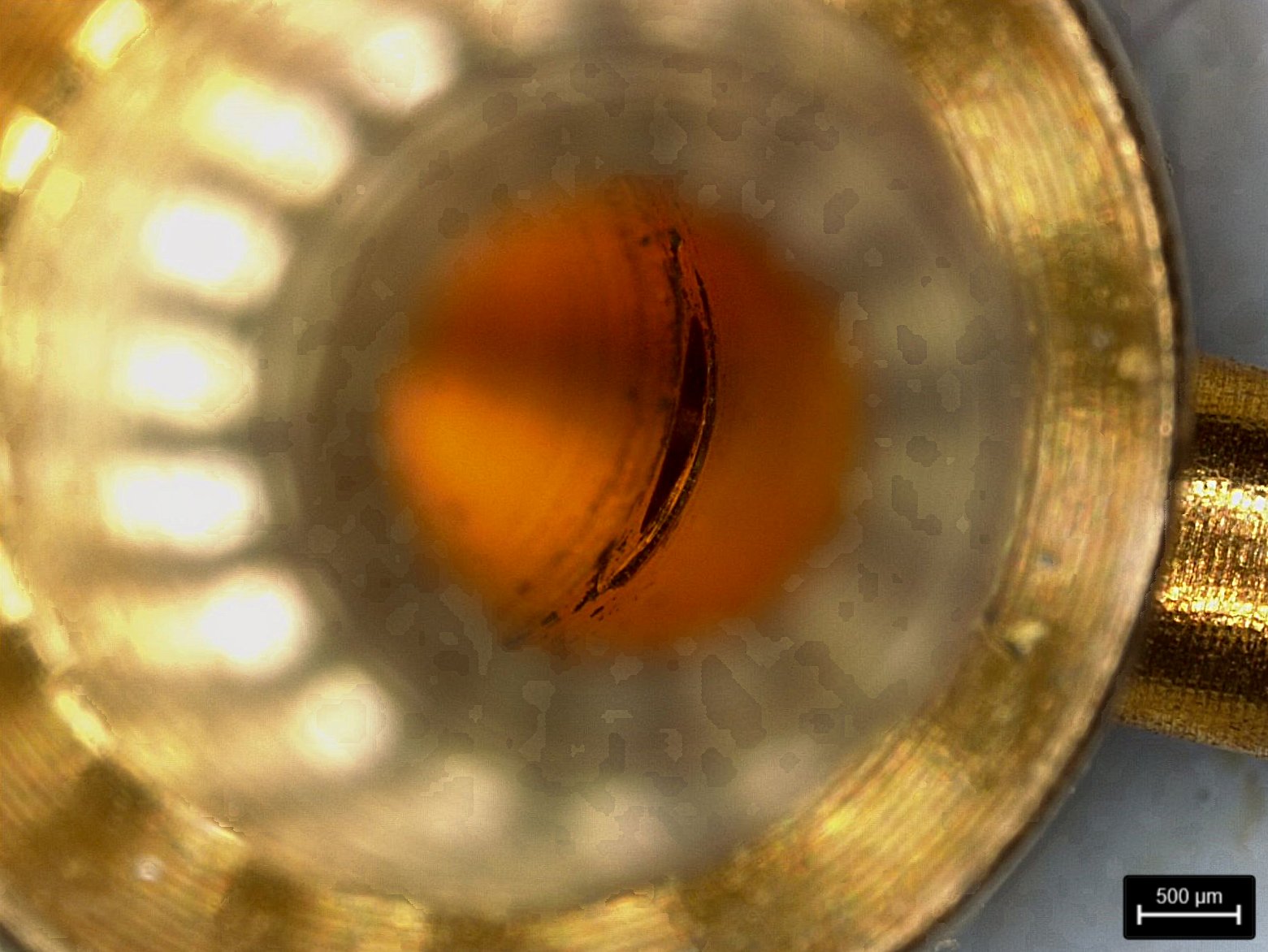}
\caption{Connection between the guidance tube and the nozzle.}
\label{fig:opt4}
\end{figure}

\begin{figure}[ht!]
\centering
\includegraphics[width=0.5\linewidth]{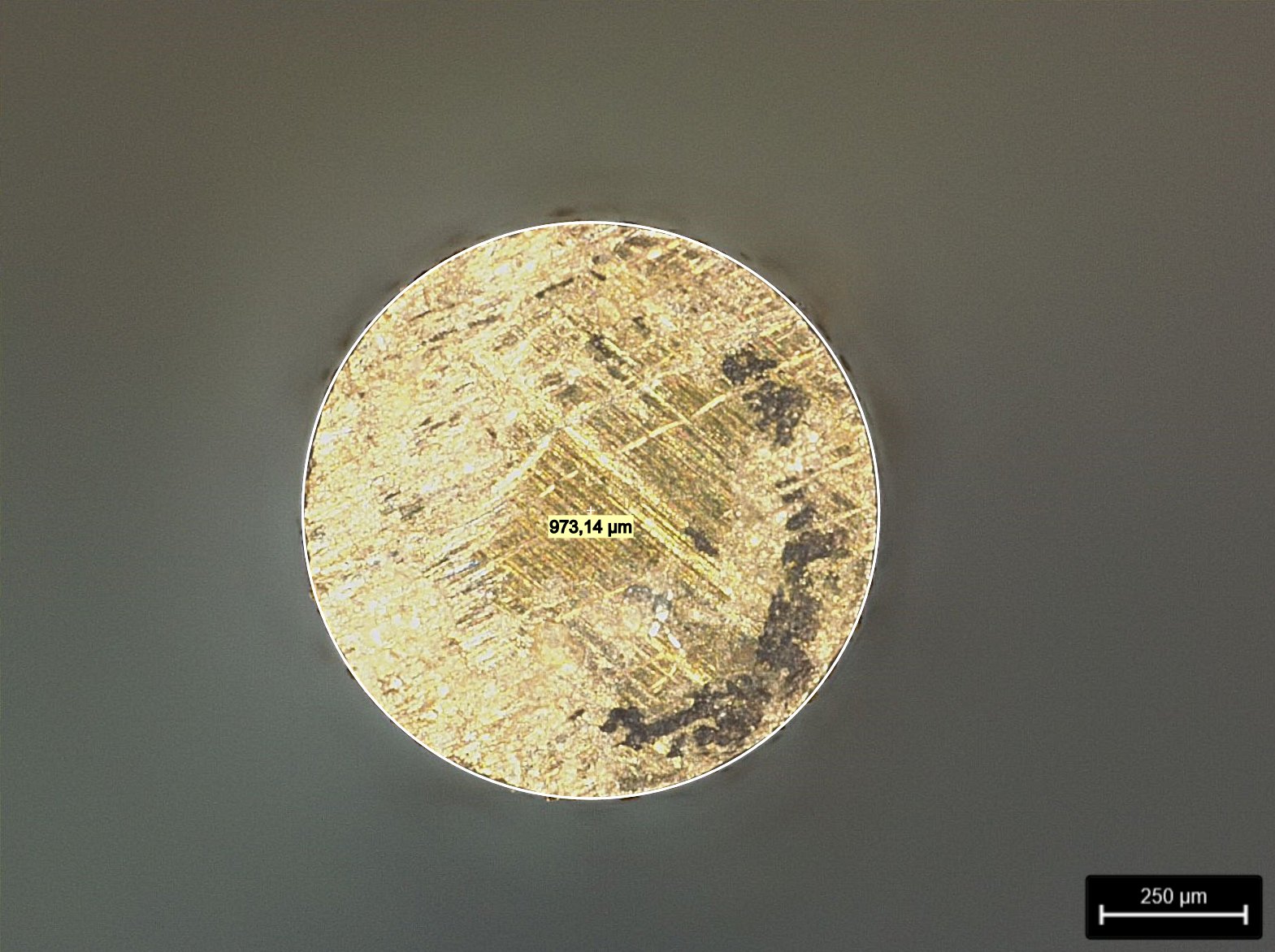}
\caption{Diameter of the pin.}
\label{fig:opt5}
\end{figure}
\clearpage

\section[\appendixname~\thesection]{Accuracy force measurement and challenges}
\label{app:ForceMes}

\begin{figure}[ht!]
\centering
\includegraphics[width=\linewidth]{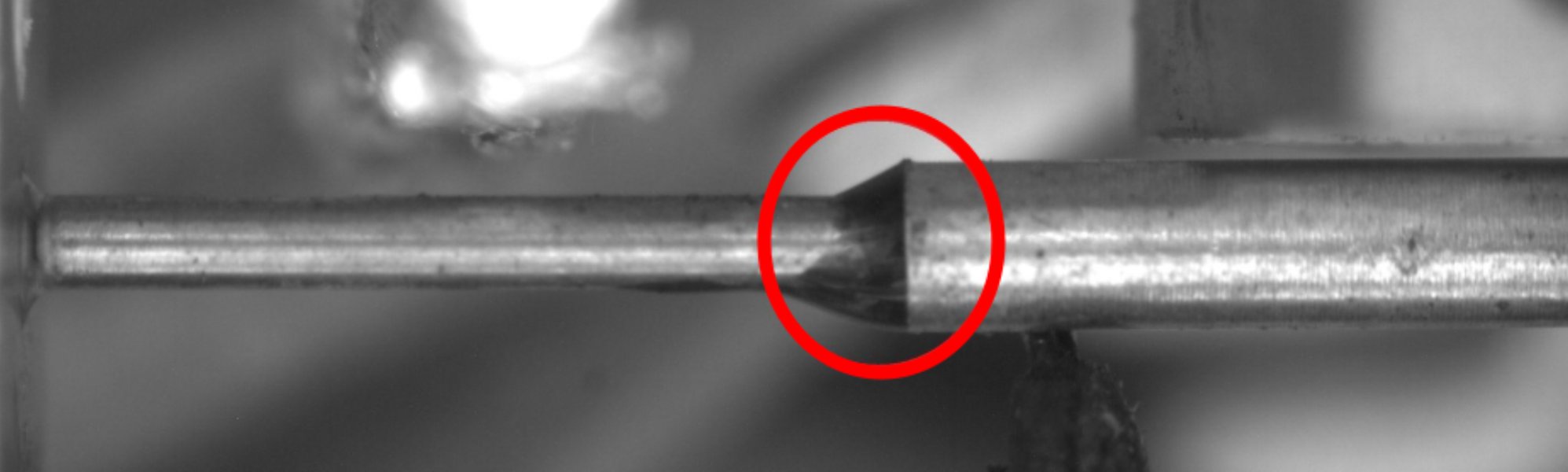}
\caption{Picture of the leakage between the guidance tube and the pin.}
\label{fig:leak}
\end{figure}

A few considerations need to be made regarding the sensorized nozzle design. \\
The polymer was recorded leaking out of the guidance tube during the experiments (Figure~\ref{fig:leak}). The effect of such leakage on the nozzle force measurement has been evaluated by imaging the exit of the guidance tube for 6.5 hours of constant extrusion. An approximation of the leakage based on those images results in a volumetric leakage of approximately 1 mm$^3$. As a comparison, when extruding at an average flow of 4 mm$^3/$s for 6.5 hours, a total volume of 93600 mm$^3$ is extruded. The flow of melt leaking out is 0.001\%. Therefore, the drag force acting on the pin, which could push it outwards (possibly increasing the recorded force value), is considered negligible. The presence of the leakage can also have a second effect on the recorded force: since the polymer has a viscoelastic behaviour, it can exert a force on the pin pulling towards the nozzle (possibly decreasing the recorded force value). We can estimate such force using the equation:
    \begin{equation}
    F = \sigma \cdot A_l   
    \end{equation}
    Where $A_l$ is the lateral area of the pin inside the tube and $\sigma$ is the stress acting on the pin. It is possible to evaluate the stress by using $\sigma = G^\prime \cdot \epsilon$, where $G^\prime$ is the elastic modulus of the leaking polymer and $\epsilon = w \cdot h$ is the strain of the leaking polymer ($w$ is the horizontal displacement of the pin and $h$ is the distance between the pin and the guidance tube). Both $h$ and $A_l$ were calculated using the length and diameter of the pin and guidance tube obtained by the microscope. Thermal expansion of the equipment was accounted for using the temperature data from the thermocouple. The friction force is different for each nozzle, flow speed and temperature. Throughout the experiments with PLA it is estimated to lie between 0\% and 5\% of the measured nozzle force. Therefore, its influence is neglected. Nozzle cleaning is quite challenging and a prolonged use of the sensorized nozzle could lead to degradation of the polymer in the tube. For this reason, during the experiment, multiple nozzles have been used.  \\

\clearpage

\section[\appendixname~\thesection]{Outputs from the \textit{streamliner} Python script}
\label{app:stream}

\begin{figure}[ht!]
\centering
\includegraphics[width=0.8\linewidth]{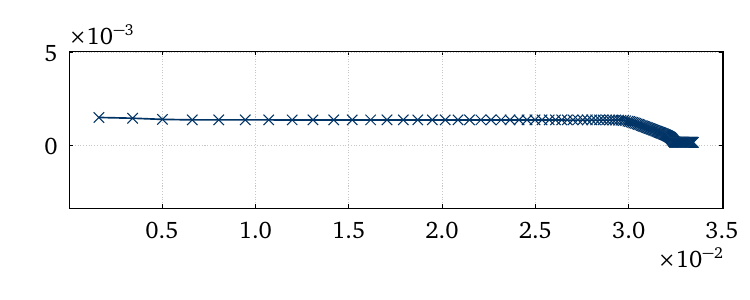}
\caption{Streamline plot to visually define the fitting points through the GUI (axes in $[m]$).}
\label{fig:str1}
\end{figure}

\begin{figure}[ht!]
\centering
\includegraphics[width=0.8\linewidth]{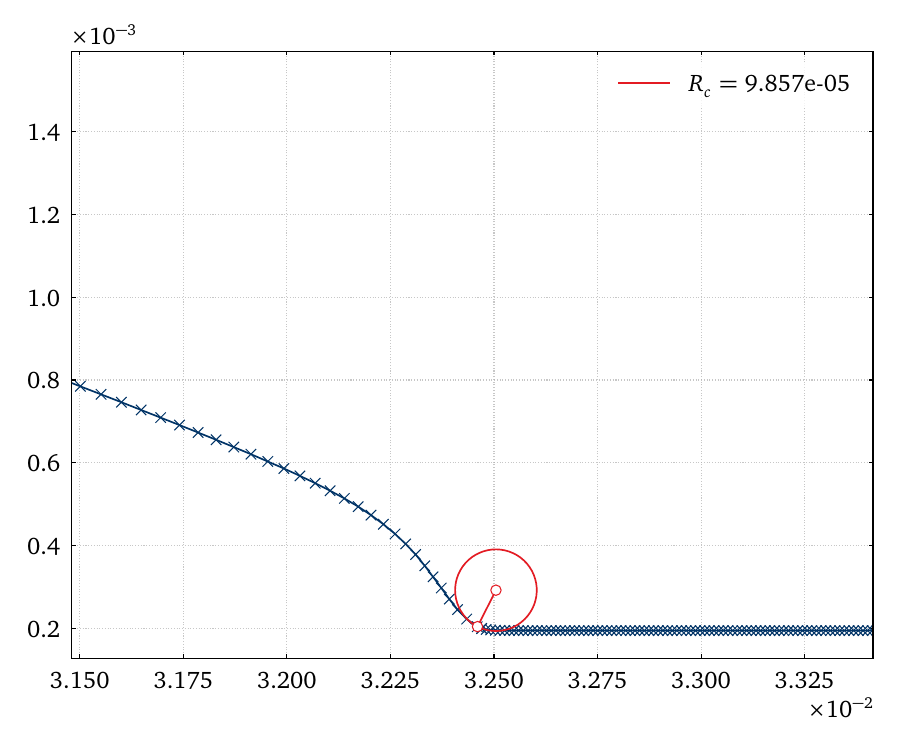}
\caption{Plotting the streamline, circumference and radius at $R_c$ measurement location (axes in $[m]$).}
\label{fig:str2}
\end{figure}

\begin{figure}[ht!]
\centering
\includegraphics[width=0.8\linewidth]{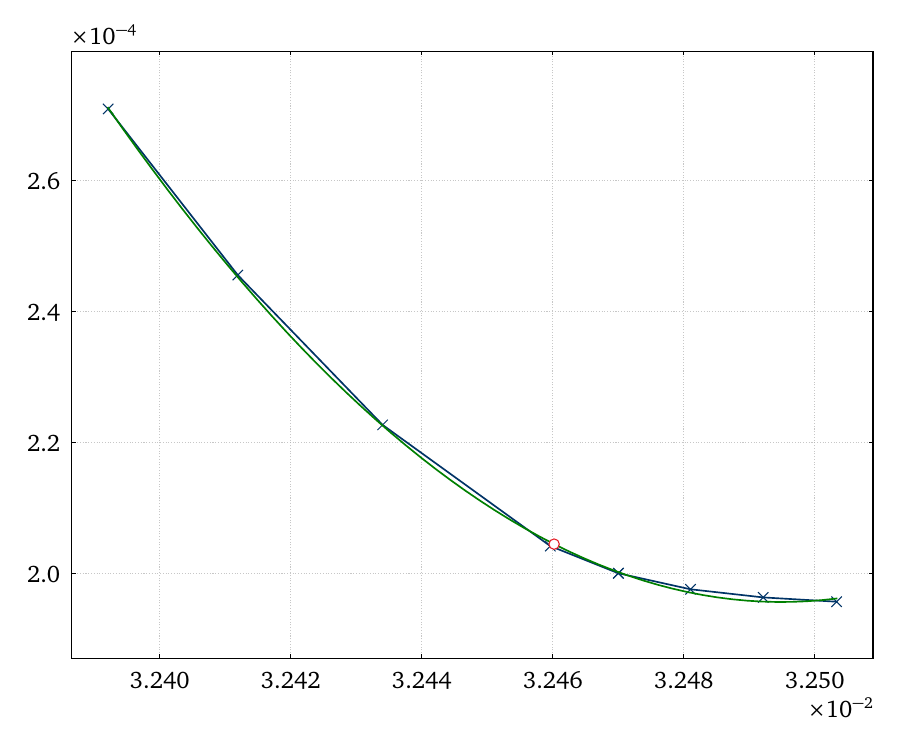}
\caption{Plotting the sliced streamline, $R_c$ location and the fitted curve (axes in $[m]$).}
\label{fig:str3}
\end{figure}

\begin{figure}[ht!]
\centering
\includegraphics[width=0.8\linewidth]{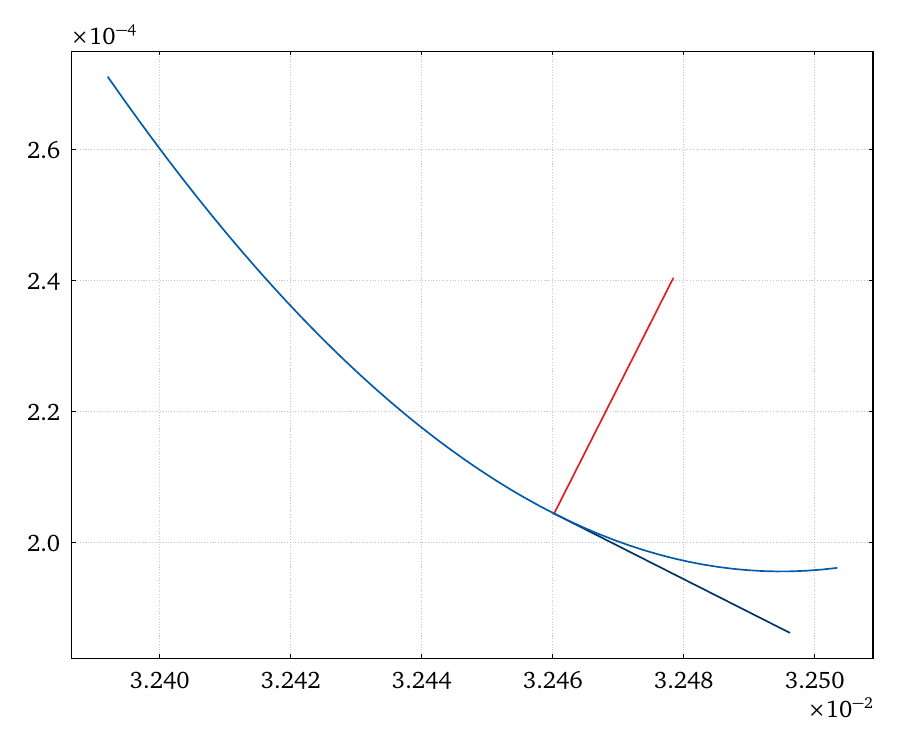}
\caption{Plotting the fitted curve, normal and tangent vectors (axes in $[m]$).}
\label{fig:str4}
\end{figure}

This section presents the output plots (Figures \ref{fig:str1} to \ref{fig:str4}) and results (Listing \ref{lst:streamliner}) from a Python script has been developed to generate, with the input of the streamline point coordinates, a polynomial fitting and closest circumference to the desired point interval and to create a new coordinate system with the normal and tangent curvature vectors, that allow for the transformation of the Cartesian tensions tensor to the new system \cite{foamscripts}. 

\clearpage

\begin{lstlisting}[caption=Terminal outputs for a streamline example.,label={lst:streamliner}]
Left limit (*1e-2)= 3.2392 <-input
Right limit (*1e-2)= 3.2516 <-input
X of point to determine Rc (*1e-2)= 3.2461 <-input
Curvature radius = 9.857e-05
Rc point coordinates =  [0.03246018 0.00020445]
X_2 vector =  [ 0.89225625 -0.45152939]
Y_2 vector =  [0.45152939 0.89225625]
Tau_xy= -362.448 <-input
Tau_xx= -166.724 <-input
Tau_yy= 225.876 <-input
Cartesian tensor matrix = 
 [[-166.724 -362.448]
 [-362.448  225.876]]
invariant1_1 = 59.152000000000015
det1 = -169027.50292799994
Transformed tensor matrix = 
 [[ 205.36485602 -372.8277361 ]
 [-372.8277361  -146.21285602]]
invariant1_2 = 59.15199999999999
det2 = -169027.50292799994
\end{lstlisting}
\clearpage

\section[\appendixname~\thesection]{Numeric/Experimental die swell comparison}
\label{app:dieswell}

\begin{figure}[ht!]
\centering
\includegraphics[width=0.4\linewidth]{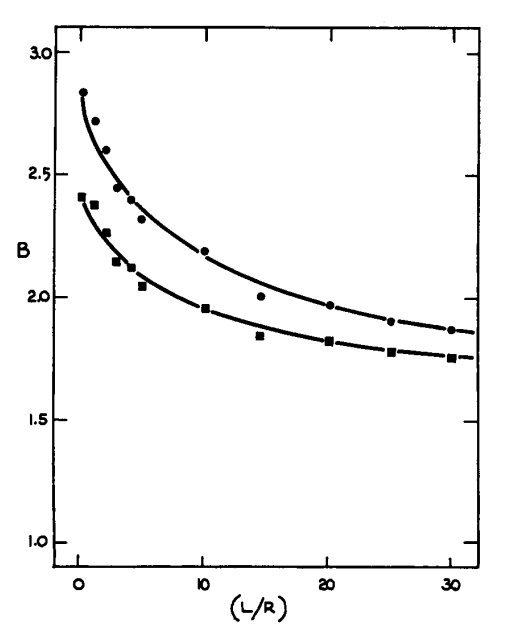}
\caption{Swelling index \textbf{B} versus capillary \textbf{L/R} for a polyethylene melt at two shear rates. With permission from \cite{Bagley1963}.}
\label{fig:bagley_ratio}
\end{figure}

Considering this, note that Equation 9 is only valid for a viscometric flow where entrance effects are negligible, as per Bagley et al. \cite{Bagley1963}. A capillary length/radius ratio of at least 15-20 is necessary for this (Figure \ref{fig:bagley_ratio}) and in this nozzle the ratio is of 5.

\begin{figure}[ht!]
\centering
\includegraphics[width=0.6\linewidth]{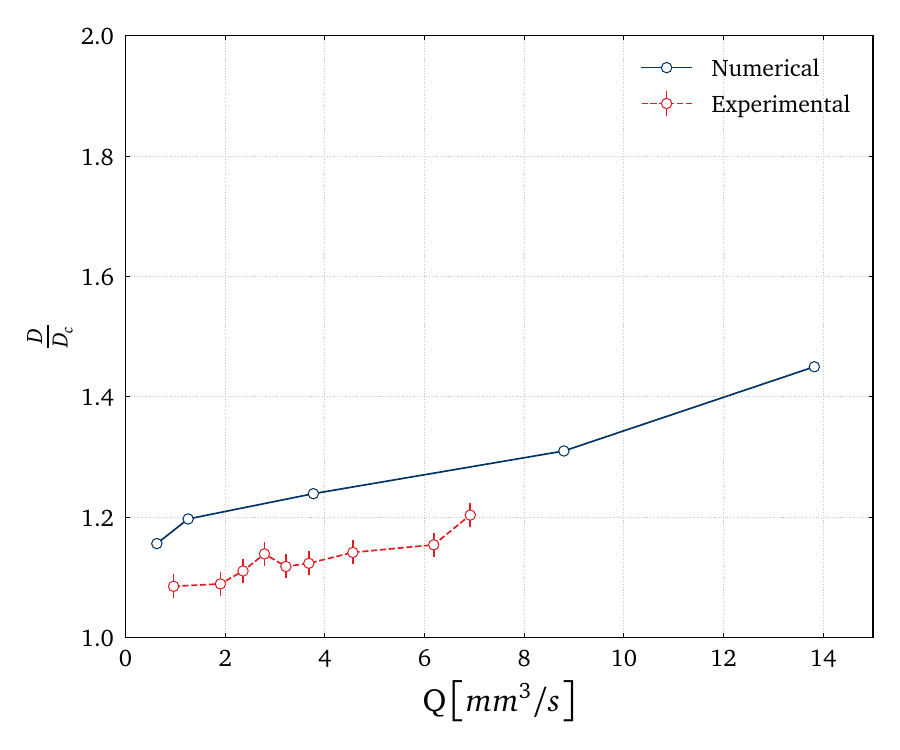}
\caption{Numerical \textit{vs} experimental die swell results - AA Core.}
\label{fig:num_exp_die_swell}
\end{figure}

Figure \ref{fig:num_exp_die_swell} compares the numerical and experimental die swell results for the AA nozzle, where the entrance effects are noticeable in the numerical plot, as also declared by \cite{Bagley1963}.

\clearpage

\section[\appendixname~\thesection]{- Experimental setup}
\label{app:imgs}

\begin{figure}[ht!]
\centering
\includegraphics[width=\linewidth]{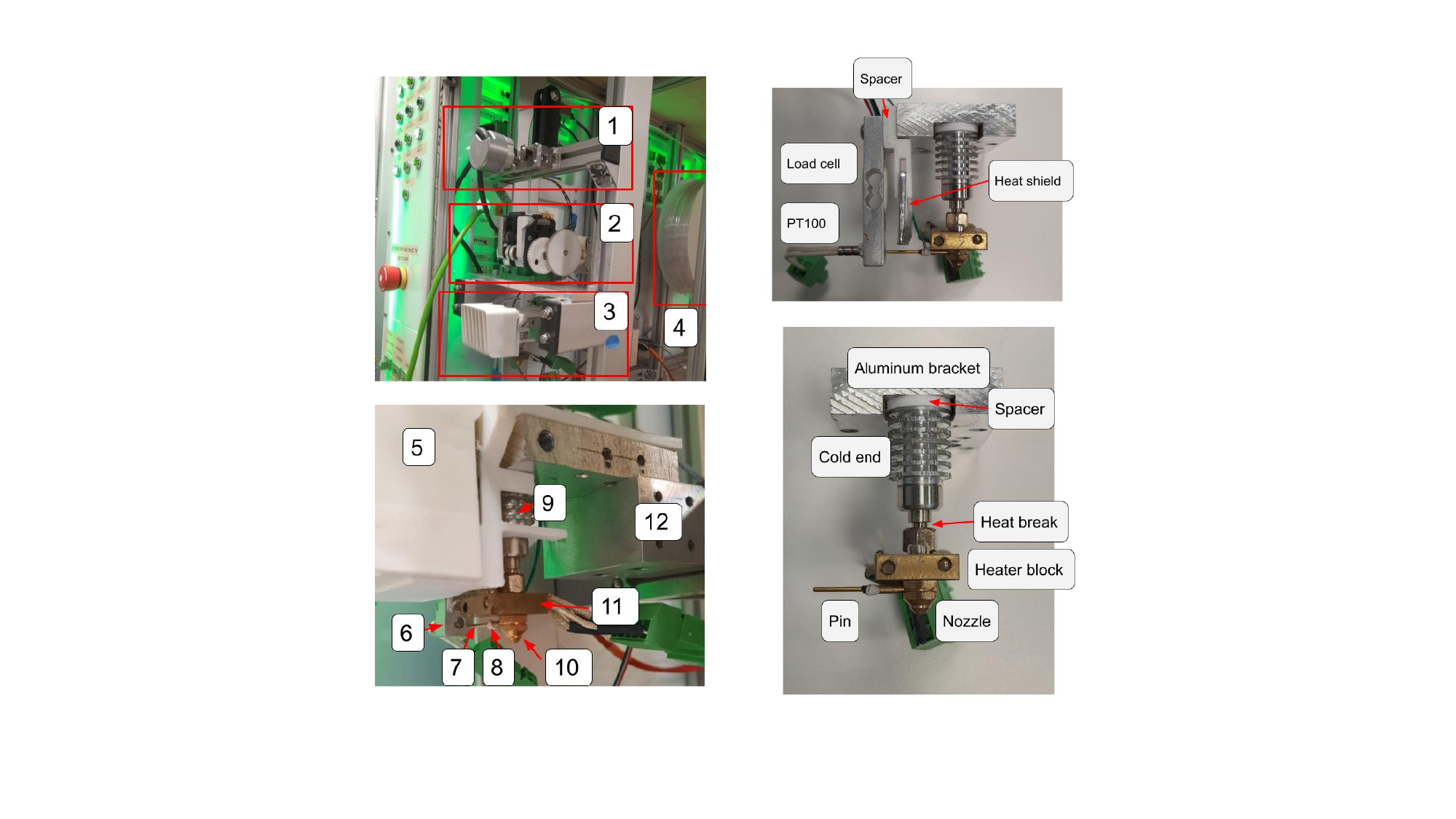}
\caption[Images of the experimental setup.]{Images of the experimental setup\protect\footnotemark.}
\label{fig:exp_setup}
\end{figure}

\footnotetext{Figure legend: 
1 - Rotary encoder; 2 - Feeder; 3 - Modified nozzle assembly; 4 - Filament;5 - Fan; 6 - 0.78 kg load cell and PT100; 7 - Pin; 8 - Thermocouple; 9 - Cold end; 10 - Nozzle; 11 - Heater and PT100; 12 - 35 kg load cell.}
 \bibliographystyle{elsarticle-num} 
 \bibliography{cas-refs}





\end{document}